\documentclass[12pt]{article}
\pdfoutput=1
\usepackage{subfigure}
\usepackage{amssymb,amsmath}
\usepackage{graphicx}
\usepackage{color}
\usepackage[colorlinks=true
,urlcolor=blue
,citecolor=blue
,linkcolor=blue
,pagecolor=blue
,linktocpage=true
,pdfproducer=medialab
]{hyperref}
\usepackage[a4paper]{geometry}

\usepackage{placeins}
\usepackage[utf8]{inputenc}
\usepackage{cancel}
\usepackage{arydshln}
\makeatletter \renewcommand{\@dotsep}{10000} \makeatother

\mathchardef\mhyphen="2D

\newcommand{\beq}{\begin{equation}}
\newcommand{\eeq}{\end{equation}}
\newcommand{\bea}{\begin{eqnarray}}
\newcommand{\eea}{\end{eqnarray}}

\newcommand{\mgut}{M_{{\rm GUT}}}

\newcommand{\WMSSM}{W_{{\rm MSSM}}}
\newcommand{\WNMSSM}{W_{{\rm NMSSM}}}

\usepackage[nodisplayskipstretch]{setspace}

\begin{document}

\begin{titlepage}
\pagestyle{empty}

\vspace*{0.2in}
\begin{center}
{\Large \bf  Charged Higgs in MSSM and Beyond
  }\\
\vspace{1cm}
{\bf  Ya\c{s}ar Hi\c{c}y\i lmaz$^{a,}$\footnote{E-mail: yasarhicyilmaz@balikesir.edu.tr}, Levent Selbuz$^{b,}$\footnote{E-mail: selbuz@eng.ankara.edu.tr}, Levent Solmaz$^{a,}$\footnote{E-mail: lsolmaz@balikesir.edu.tr} and Cem Salih \"{U}n$^{c,}$\footnote{E-mail: cemsalihun@uludag.edu.tr}}
\vspace{0.5cm}
\begin{flushleft}
{\it $^a$Department of Physics, Bal\i kesir University, TR10145, Bal\i kesir, Turkey} \\
{\it $^b$Department of Engineering Physics, Ankara University, TR06100, Ankara, Turkey} \\
{\it $^c$Department of Physics, Uluda\~{g} University, 16059, Bursa, Turkey}
\end{flushleft}

\end{center}

\vspace{0.5cm}
\begin{abstract}
\noindent We conduct a numerical study over the constrained MSSM (CMSSM), Next-to-MSSM (NMSSM) and $U(1)$ extended MSSM (UMSSM) to probe the  allowed mass ranges of the charged Higgs boson and its dominant decay patterns, which might come into prominence in near future collider experiments. {We present results obtained from a limited scan for CMSSM} as a basis and compare its predictions with the extended models. We observe within our data that a wide mass range is allowed as $0.5(1) \lesssim m_{H^{\pm}} \lesssim 17$ TeV in UMSSM (NMSSM). We find that the dominant decay channel is mostly $H^{\pm}\rightarrow tb$ such that ${\rm BR}(H^{\pm}\rightarrow tb) \sim 80\%$. While this mode remains dominant over the whole allowed parameter space of CMSSM, we realize some special domains in the NMSSM and UMSSM, in which ${\rm BR}(H^{\pm}\rightarrow tb) \lesssim 10\%$. In this context, the decay patterns of the charged Higgs can play a significant role to distinguish among the SUSY models. In addition to the $tb$ decay mode, we find that the narrow mass scale in CMSSM allows only the decay modes for the charged Higgs boson to $\tau\nu$ ($\sim 16\%$), and their supersymmetric partners $\tilde{\tau}\tilde{\nu}$ ($\sim 13\%$). On the other hand, it is possible to realize the mode in NMSSM and UMSSM in which the charged Higgs boson decays into a chargino and neutralino pair up to about $25\%$. {This decay mode requires non-universal boundary conditions within the MSSM framework to be available, since CMSSM yields ${\rm BR} (H^{\pm}\rightarrow \tilde{\chi}_{1}^{0}\tilde{\chi}_{1}^{\pm}) \lesssim 1\%$.} It can also be probed in near future collider experiments through the missing energy and CP-violation measurements.  Moreover, the chargino mass is realized as $m_{\tilde{\chi}_{1}^{\pm}} \gtrsim 1$ TeV in NMSSM and UMSSM, and these solutions will be likely tested soon in collider experiments through the chargino-neutralino production. Focusing on the chargino-neutralino decay patterns, we also present tables which list the possible ranges for the charged Higgs production and decay modes.

\end{abstract}
\end{titlepage}

\section{Introduction}
\label{sec:intro}

After the null results for the new physics, the current experiments
have focused on the Higgs boson properties and analyses have been
quite enlarged such that the Higgs boson couplings and decays are
now being studied precisely. The Higgs boson itself is a strong hint
for the new physics and there are some drawbacks of the Standard
Model (SM) such as the gauge hierarchy problem
\cite{Gildener:1976ai} and the absolute stability of the SM Higgs
potential \cite{hinstability}. In addition, most of the models
beyond the SM (BSM) need to enlarge the Higgs sector so that their
low scale phenomenology includes extra Higgs bosons which are not
present in the SM. Among many well motivated BSM models,
supersymmetric models take arguably a special place, since they are
able to solve the gauge hierarchy problem, provide plausible
candidates for the dark matter (DM) and so on. Besides, the Higgs
sector in such models requires two Higgs doublets and so the low
scale Higgs sector includes two CP-even Higgs bosons ($h,H$), one
CP-odd Higgs boson ($A$) and two charged Higgs bosons ($H^{\pm}$).

While the lightest CP-even Higgs boson is expected to exhibit very
similar properties to the SM-like Higgs boson at the decoupling
limit ($m_{H}\sim m_{A}\sim m_{H^{\pm}}\gg m_{h}$), its couplings to
the SM particles can still deviate from the SM and thus, it can be
constrained by such deviations \cite{Haber:1993pv}. Similarly, the
heavier Higgs bosons can be constrained if they can significantly
decay into the SM particles. For instance, if the CP-odd Higgs boson
decays mostly into a pair of $\tau-$leptons, then its mass can be
constrained as $m_{A}\gtrsim 1$ TeV depending on $\tan\beta$
\cite{Khachatryan:2014wca}. In addition, the flavor changing decays
of $B-$meson also yield important implications for these Higgs
bosons. The $B_{s}\rightarrow \mu^{+}\mu^{-}$ process receives some
contributions from $A-$boson proportional to
$(\tan\beta)^{6}/m_{A}^{4}$ \cite{Choudhury:1998ze}, and the strong
agreement between the experimental results (${\rm
BR}(B_{s}\rightarrow \mu^{+}\mu^{-})=3.2^{+1.5}_{-1.2}\times
10^{-9}$ \cite{Aaij:2012nna}) and the SM (${\rm BR}(B_{s}\rightarrow
\mu^{+}\mu^{-})=(3.2\pm 0.2)\times 10^{-9}$ \cite{Buchalla:1995vs})
strictly constrains the $A-$boson phenomenology.

Among these extra Higgs bosons, the charged Higgs boson plays a
crucial role, since the SM does not have any charged scalar. It can
be produced at the current collider experiments along with other
particles as $pp\rightarrow (t,W^{\pm},t\bar{b},\ldots)H^{\pm}$,
where $p$ stands for the proton. Even though its production is
rather difficult and production cross-section is small in comparison
to the other particles, the charged Higgs boson can be {visible}
with large center of mass energy and luminosity in near future
collider experiments. In this context, the track of its decays can
be directly related to the new physics. Furthermore, this
distinguishing charged particle may reveal itself in many
manifestations and different supersymmetric models may favor
different predictions. Even  though the usual dominant decay mode is
$H^{\pm}\rightarrow t\bar{b}$ when $m_{H^{\pm}}\gtrsim m_{t}+m_{b}$
in the CMSSM,  models with extended  particle content and/or gauge
group  can open window for other  probable and important decay
modes. In addition, richer phenomenology can be revealed when the
charged Higgs boson is allowed to decay into new supersymmetric
particles. For instance, if the $H^{\pm}\rightarrow
\tilde{\chi}_{i}^{0}\tilde{\chi}_{j}^{\pm}$ mode is open, one can
also measure the CP-asymmetry throughout such processes
\cite{Frank:2007zza}.

Based on different decay patterns of the charged Higgs boson, we
analyzed the charged Higgs boson in this work within three different models which are the
minimal supersymmetric extension of the SM (MSSM), $U(1)$ extended
MSSM (UMSSM) and Next-to-MSSM (NMSSM). We restrict our analyses to
constrained versions of these models, in which the low scale
observables can be determined with only a few input parameters defined at
the grand unification scale ($\mgut$). Of course, a more detailed
study can be performed from weak scale side which is a tedious work
and this is beyond the scope of this paper. Throughout the analyses
the CMSSM framework will be considered as the base and implications
of the other two models will be discussed in a way that also
compares them with CMSSM. It is important to stress that the
different imprints of the charged Higgs boson can be used to
distinguish a model from  another and this may be useful for future
charged Higgs studies.

The outline of the rest of the paper is the following: We will
briefly describe the models under concern in Section
\ref{sec:model}. After we summarize the scanning procedure, employed
experimental constraints in Section
\ref{sec:scan}, Section \ref{sec:spectrum} discusses the
mass spectrum in terms of the particles, which can participate in
the charged Higgs boson decay modes. The results for the production
and decay modes of the charged Higgs boson are {presented} in
Section \ref{sec:res}. We also present tables containing rates for
the charged Higgs boson production and its decay modes over some
benchmark points in this section. Section \ref{sec:conc} summarizes
and concludes our findings.

\section{Models}
\label{sec:model}

\subsection{MSSM}
\label{subsec:MSSM}

The superpotential in MSSM is given as

\begin{equation}
\WMSSM=\mu \hat{H}_{u}\hat{H}_{d}+Y_{u} \hat{Q} \hat{H}_{u} \hat{U}+Y_{d}\hat{Q}\hat{H}_{d} \hat{D}+Y_{e}\hat{L}\hat{H}_{d}\hat{E}
\label{MSSMW}
\end{equation}
where $\mu$ is the bilinear mixing term for the MSSM Higgs doublets
$H_{u}$ and $H_{d}$; $Q$ and $L$ denote the left handed squark and
lepton dublets, while $U,D,E$ stand for the right-handed u-type
squarks, d-type squarks and sleptons respectively. $Y_{u,d,e}$ are
the Yukawa couplings between the Higgs fields and the matter fields
shown as subscripts. {Higgsino mass term} $\mu$ is included in the SUSY preserving
Lagrangian in MSSM, and hence it is allowed to be at any scale from
the electroweak (EW) scale to $\mgut$. In this sense, even though it
is relevant to the EW symmetry breaking, its scale is not
constrained by the EW symmetry breaking scale ($\sim 100$ GeV). This
is called the $\mu-$problem in MSSM. In addition to $\WMSSM$, the
soft SUSY breaking (SSB) Lagrangian is given below

\begin{equation*}\hspace{-5.7cm}
-\mathcal{L}^{\cancel{SUSY}}_{{\rm MSSM}}=m^{2}_{H_{u}}|H_{u}|^{2}+m^{2}_{H_{d}}|H_{d}|^{2}+m^{2}_{\tilde{Q}}|\tilde{Q}|^{2}+m^{2}_{\tilde{L}}|\tilde{L}|^{2}
\end{equation*}
\begin{equation*}
+m^{2}_{\tilde{U}}|\tilde{U}|^{2}+m^{2}_{\tilde{D}}|\tilde{D}|^{2}+m^{2}_{\tilde{E}}|\tilde{E}|^{2}+\sum_{a}M_{a}\lambda_{a}\lambda_{a}+(B\mu H_{u}H_{d}+{\rm h.c.})
\end{equation*}
\begin{equation}\hspace{-3.7cm}
+A_{u}Y_{u}\tilde{Q}H_{u}\tilde{U^{c}}+A_{d}Y_{d}\tilde{Q}H_{d}\tilde{D^{c}}+A_{e}Y_{e}\tilde{L}H_{d}\tilde{E^{c}}
\label{MSSM_SSB}
\end{equation}
where the field notation is as given before. In addition,
$m_{\phi}^{2}$ with
$\phi=H_{u},H_{d},\tilde{Q},\tilde{L},\tilde{U},\tilde{D},\tilde{E}$
are the SSB mass terms for the scalar fields. $A_{u,d,e}$ are the
SSB terms for the trilinear scalar interactions, while $B$ is the
SSB bilinear mixing term for the MSSM Higgs fields. After adding the
SSB Lagrangian, the Higgs potential in MSSM becomes more complicated
than the SM, and the masses of the physical Higgs bosons can be
found in terms of $\mu$, $m_{H_{u}}$, $m_{H_{d}}$ and $\tan\beta$,
where $\tan\beta = v_{u}/v_{d}$ is the ratio of the vacuum
expectation values (VEVs) of the MSSM Higgs fields. The tree level
Higgs boson masses can be found as \cite{Martin:1997ns};

\begin{equation}
\begin{array}{ll}
m_{h,H} & = \dfrac{1}{2}\left( m_{A}^{2}+M_{Z}^{2}\mp \sqrt{(m_{A}^{2}-M_{Z}^{2})^{2}+4M_{Z}^{2}m_{A}^{2}\sin^{2}(2\beta)}   \right) \\ & \\
m_{H^{\pm}}^{2} & =m_{A}^{2}+M_{W}^{2}  \\ & \\
m_{A}^{2} & =2|\mu|^{2}+m_{H_{u}}^{2}+m_{H_{d}}^{2}
\end{array}
\label{HMtree}
\end{equation}
where $M_{Z}$ and $M_{W}$ are the masses of the $Z-$ and $W-$bosons
respectively. As it is well-known, the lightest CP-even Higgs boson
tree level mass is problematic in MSSM, since it is bounded by
$M_{Z}$ from above as $m_{h}^{2} \lesssim
M_{Z}^{2}\cos^{2}(2\beta)$. This conflict can be resolved by adding
the loop corrections to the Higgs boson mass. Utilizing the loop
corrections to realize 125 GeV Higgs boson at the low scale requires
either multi-TeV stop mass or relatively large SSB trilinear
$A-$term \cite{Heinemeyer:2011aa}. In this context the 125 GeV Higgs
boson constraint leads to heavy spectrum in SUSY particles
especially in the CMSSM framework where all scalar
masses are set by a single parameter at $\mgut$. Besides, if the
mass scales for the extra Higgs bosons are realized as $m_{A}\sim
m_{H}\sim m_{H^{\pm}} \gtrsim 1$ TeV, it requires large $m_{H_{u}}$
and $m_{H_{d}}$ as seen from Eqs.(\ref{HMtree}). It also brings the
naturalness problem back to the SUSY models, since the consistent
electroweak symmetry breaking scale requires $\mu \approx m_{H_{u}}$
over most of the fundamental parameter space of the models. It also
arises the $\mu-$problem in MSSM mentioned above.

The $\mu-$term is also important since it is the Higgsino masses at
the low scale. In this context, if $\mu$ term is significantly low
in comparison to the gaugino masses $M_{1}$ and $M_{2}$, the LSP
neutralino can exhibit Higgsino-like properties, and it yields
different DM phenomenology. Nature of the DM can be investigated by
considering the neutralino mass matrix given as

\begin{equation}
\mathcal{M}_{\tilde{\chi}^{0}}^{{\rm MSSM}}=\left(\begin{array}{cccc} M_{1} & 0 & -\dfrac{g_{1}v_{d}}{\sqrt{2}} & \dfrac{g_{1}v_{u}}{\sqrt{2}}  \\ &&&\\
0 & M_{2} & \dfrac{g_{2}v_{d}}{\sqrt{2}} & -\dfrac{g_{2}v_{u}}{\sqrt{2}} \\ &&& \\
-\dfrac{g_{1}v_{d}}{\sqrt{2}} & \dfrac{g_{2}v_{d}}{\sqrt{2}} & 0 & -\mu \\&&&\\
\dfrac{g_{1}v_{u}}{\sqrt{2}} & -\dfrac{g_{2}v_{u}}{\sqrt{2}} & -\mu & 0
 \end{array}\right)
 \label{Chi_MSSM}
\end{equation}
in the basis
$(\tilde{B},\tilde{W},\tilde{h}_{d},\tilde{h}_{u})$, where
$\tilde{B}$ and $\tilde{W}$ denote Bino and Wino respectively, which
may be called electroweakinos, while $\tilde{h}_{d}$ and
$\tilde{h}_{u}$ represent the Higgsinos from $H_{d}$ and $H_{u}$
superfields respectively. Similarly, the chargino mass matrix can be
written as

\begin{equation}
\mathcal{M}_{\tilde{\chi}^{\pm}}=\left(\begin{array}{cc}
M_{2} & \dfrac{1}{\sqrt{2}}g_{2}v_{u} \\ & \\ \dfrac{1}{\sqrt{2}}g_{2}v_{d} & \mu
\end{array}\right).
\label{chargino_MSSM}
\end{equation}

The properties and relevant phenomenology involving with the
chargino and neutralino can be understood by comparing $M_{1}$,
$M_{2}$ and $\mu$. If $\mu > M_{1},M_{2}$ then both LSP neutralino
and the lightest chargino exhibit gaugino properties and the gauge
couplings are dominant in strength of the relevant interactions.
When $\mu < M_{1},M_{2}$, the LSP neutralino and the lightest
chargino are mostly formed by the Higgsinos, and the Yukawa
couplings also take part in interactions as well as the gauge
couplings. Since, the charged Higgs is, in principle, allowed to
decay into a pair of a neutralino and a chargino, the self
interaction couplings in the Higgs sector are also important in such
decay processes, when the chargino and neutralino are Higgsino-like.

\subsection{NMSSM}
\label{subsec:NMSSM}

The main idea behind NMSSM is to resolve the $\mu-$problem of MSSM
in a dynamic way that an additional $S$ field generates the
$\mu-$term by developing a non-zero VEV. It can be realized by
replacing the term with $\mu-$term in $\WMSSM$ given in
Eq.(\ref{MSSMW}) with a trilinear term as $h_{s}SH_{u}H_{d}$, where
$S$ is chosen preferably to be singlet under the MSSM gauge group.
If there is no additional term depending on the $S-$field, then the
Lagrangian also remains invariant under Pecce-Quinn (PQ) like
symmetry which transforms the fields as follows

\begin{equation}
\begin{array}{ccc}
H_{u}\rightarrow e^{iq_{PQ}\theta}H_{u} & H_{d}\rightarrow e^{iq_{PQ}\theta}H_{d} & S\rightarrow e^{-2iq_{PQ}\theta} \\ & & \\
Q\rightarrow e^{-iq_{PQ}\theta}Q & L\rightarrow e^{-iq_{PQ}\theta}L & U,D,E \rightarrow U,D,E
\end{array}
\label{PQsymm}
\end{equation}

Such a symmetry in the Lagrangian can help to resolve the strong
CP-problem \cite{Peccei:1977ur}. However, in NMSSM, the Pecce-Quinn
symmetry is broken by the $\mu-$term spontaneously, since it happens
by the VEV of the $S$ field. In this case, there has to be a
massless Goldstone boson, which can be identified as axion. Such a
massless field is strongly constrained by the cosmological
observations \cite{Hagiwara:2002fs}. Moreover, $h_{s}$ is restricted
into a very narrow range ($10^{-10} \gtrsim h_{s} \lesssim 10^{-7}$)
experimentally, and in order to yield $\mu \sim \mathcal{O}(100)$
GeV VEV of S should be very large, and it brings back the
naturalness problem.

The situation of the massless Goldstone boson arose from the
spontaneous breaking of the PQ symmetry, is to add another $S$
dependent term as $\dfrac{1}{3}\kappa S^{3}$, which explicitly
breaks the PQ symmetry, while $\mathbb{Z}_{3}$ symmetry remains
unbroken. However, despite avoiding the massless axion in this case,
the effectively generated $\mu-$term spontaneously breaks the
discrete $\mathbb{Z}_{3}$ symmetry, which arises the domain-wall
problem. This problem can be resolved by adding non-renormalizable
higher order operators which break $\mathbb{Z}_{3}$ symmetry, while
preserving the $\mathbb{Z}_{2}$ symmetry at the Planck scale. (For a
detailed description of the domain-wall problem and its resolution,
see \cite{Panagiotakopoulos:1998yw}). In our work we assume that the
domain-wall problem is resolved and we consider the following
superpotential in the NMSSM framework;

\begin{equation}
\WNMSSM=\WMSSM(\mu =0)+h_{s}\hat{S}\hat{H}_u \hat{H}_d +\dfrac{1}{3}\kappa \hat{S}^{3}
\label{NMSSMW}
\end{equation}
and the corresponding SSB Lagrangian is
\begin{equation}
\mathcal{L}_{{\rm NMSSM}}^{\cancel{SUSY}}=\mathcal{L}_{{\rm MSSM}}^{\cancel{SUSY}}(\mu=0)-m_{S}^{2}S^{*}S-\left[h_{s}A_{s}SH_{u}H_{d}+\dfrac{1}{3}\kappa A_{\kappa}S^{3}+{\rm h.c.} \right]
\label{NMSSM_SSB}
\end{equation}
where $\mathcal{L}_{{\rm MSSM}}^{\cancel{SUSY}}(\mu=0)$ is the SSB
Lagrangian for MSSM given in Eq.(\ref{MSSM_SSB}) with $\mu=0$,
$m_{S}$ is the mass for the scalar component of $S$, while $A_{s}$
and $A_{\kappa}$ are trilinear SSB terms for the scalar
interactions. According to the superpotential and the SSB
Lagrangian, the second term in Eq.(\ref{NMSSMW}) is responsible for
generating the $\mu-$term effectively as $\mu_{eff}\equiv
h_{s}v_{s}/\sqrt{2}$ and the first term in the paranthesis of
Eq.(\ref{NMSSM_SSB}) is $B\mu$ correspondence. Although the particle
content is simply enlarged by including an extra singlet field in
NMSSM, the neutral scalar component of this field can mix with the
CP-even and CP-odd Higgs boson of MSSM, while the charged Higgs
sector remains intact. After the EW symmetry breaking, the NMSSM
Higgs sector includes three CP-even Higgs bosons, two CP-odd Higgs
bosons and two charged Higgs bosons. A detailed discussion for the
Higgs sector can be found in Ref. \cite{Ellwanger:2009dp}. Once the
mass matrix for the CP-even Higgs boson states is diagonalized, the
lightest mass eigenvalue can be found as

\begin{equation}
m_{h}^{2}=M_{Z}\left(\cos^{2}(2\beta) + \dfrac{h_{s}}{g}\right)
\end{equation}
where the first term covers the MSSM part of the Higgs boson, while
the second term encodes the contributions to the tree-level Higgs
boson mass from the singlet. In this sense, the necessary loop
corrections to the Higgs boson mass may be relaxed and lighter mass
spectrum for the SUSY particles can be realized. As the  lightest
CP-even Higgs boson, other Higgs bosons receive contributions from
the singlet, and the tree-level mass for the charged Higgs boson can
be obtained as \cite{Drees:1988fc}

\begin{equation}
m^{2}_{H^{\pm}}=M_{W}^{2}+\dfrac{2h_{s}v_{s}}{\sin(2\beta)}(A_{s}+\kappa v_{s})-h_{s}(v_{u}^{2}+v_{d}^{2})
\end{equation}

In addition to the Higgs sector, the neutralino sector of NMSSM has
five neutralinos including to the supersymmetric partner of $S$
field - so-called singlino - in addition to the MSSM neutralinos.
The mass matrix for the neutralinos {in} the basis ($\tilde{B}$,
$\tilde{W}$, $\tilde{H}_{u}$, $\tilde{H}_{d}$, $\tilde{S}$) is
obtained as

\begin{equation}
\mathcal{M}_{\tilde{\chi}^{0}}^{\rm NMSSM}=\left(\begin{array}{cccc:c} &&&& 0 \\ &&&& 0 \\
\multicolumn{4}{c:}{\mathcal{M}_{\tilde{\chi}^{0}}^{\rm MSSM}(\mu=\mu_{eff})} & -h_{s}v_{u} \\ &&&& -h_{s}v_{d} \\  \hdashline
&&&&\\
0 & 0 & -h_{s}v_{u} & -h_{s}v_{d} & 2\kappa v_{s}
 \end{array}\right)
 \label{Chi_NMSSM}
\end{equation}
where $\mathcal{M}_{\tilde{\chi}^{0}}^{\rm MSSM}(\mu=\mu_{eff})$ is
the MSSM neutralino masses and mixing as given in
Eq.(\ref{Chi_MSSM}), while the extra column and row represent the
mixing of the singlino with the MSSM neutralinos. As seen from the
$\mathcal{M}_{\tilde{\chi}^{0}}^{55}$, the singlino mass is found as
$M_{\tilde{S}}=2\kappa v_{s}$. Even though it is left out from the
first $4\times 4$ matrix of the MSSM neutralinos, the singlet sector
is still effective, since the mass term for the Higgsinos is
determined by the VEV of $S$ field as
$\mu_{eff}=h_{s}v_{s}/\sqrt{2}$. If one assumes the lightest mass
eigenvalue of $\mathcal{M}_{\tilde{\chi}^{0}}$ is also the lightest
supersymmetric particle (LSP), species of the LSP can yield
dramatically different dark matter (DM) implications than those
obtained in the MSSM framework, especially when the singlino is
realized so light that it can significantly mix with the other
neutralinos in formation of the LSP neutralino. The chargino sector
remains intact, since NMSSM does not introduce any new charged field to the
particle content; hence, the chargino masses and mixing are given
with the same matrix given in Eq.(\ref{chargino_MSSM}). Note that
VEV of the singlet field is indirectly effective in the chargino
sector through the Higgsino mass, which can be seen by replacing
$\mu$ with $\mu_{eff}$ in Eq.(\ref{chargino_MSSM}).

\subsection{UMSSM}
\label{subsec:UMSSM}

In the previous subseciton, where we discussed NMSSM, even though
its non-zero VEV might be expected to break a gauged symmetry
spontaneously, there was no symmetry whose breaking is triggered
with VEV of $S$ except the global PQ and $\mathbb{Z}_{3}$
symmetries. In this sense, one can associate a gauged symmetry to
$v_{s}$ by extending the MSSM gauge group with a simple abelian
$U(1)'$ symmetry \cite{Demir:2005ti,Langacker:1998tc}. Such
extensions of MSSM form a class of $U(1)'$ models (UMSSM), and its
gauge structure can be originated to the GUT scale, when the
underlying symmetry group at $\mgut$ is larger than $SU(5)$. The
most interesting breaking pattern, which results in UMSSM, can be
realized when the GUT symmetry is identified with the exceptional
group $E_{6}$:

\begin{equation}
E_{6}\rightarrow SO(10)\times U(1)_{\psi}\rightarrow SU(5)\times U(1)_{\chi}\times U(1)_{\psi}\rightarrow G_{{\rm MSSM}}\times U(1)'
\label{E6breaking}
\end{equation}
where $G_{{\rm MSSM}}=SU(3)_{c}\times SU(2)_{L}\times U(1)_{Y}$ is
the MSSM gauge group, and $U(1)'$ can be expressed as a general
mixing of $U(1)_{\psi}$ and $U(1)_{\chi}$ as

\begin{equation}
U(1)'=\cos \theta_{E_{6}}U(1)_{\chi}+\sin\theta_{E_{6}}U(1)_{\psi}.
\label{Umixing}
\end{equation}

In a general treatment, all the fields including the MSSM ones are
allowed to have non-zero charges under the $U(1)'$ gauge group;
thus, despite the similarity with NMSSM, the invariance under the
$U(1)'$ symmetry does not allow the term $\kappa S^{3}$ as well as
$\mu H_{u}H_{d}$. The charge configurations of the fields for
$U(1)_{\psi}$ and $U(1)_{\chi}$ models are given in Table
\ref{charges}. The charge configuration for any $U(1)'$ model can be
obtained with the mixing of $U(1)_{\psi}$ and $U(1)_{\chi}$, which
is quantified with the mixing angle $\theta_{E_{6}}$, through the
equation provided below Table \ref{charges}.

\begin{table}[ht!]
\setstretch{1.5}
\centering
\begin{tabular}{|c|c|c|c|c|c|c|c|c|c|c|}
\hline
 Model & $\hat{Q}$ & $\hat{U}^{c}$ & $\hat{D}^{c}$ & $\hat{L}$ & $\hat{E}^{c}$ & $\hat{H}_{d}$ & $\hat{H}_{u}$ & $\hat{S}$ \\
 \hline
$ 2\sqrt{6}~U(1)_{\psi}$ & 1 & 1 & 1 &1 &1 & -2 & -2 & 4\\
\hline
$ 2\sqrt{10}~U(1)_{\chi}$ & -1 & -1 & 3 & 3 & -1 & -2 & 2 & 0 \\
\hline
\end{tabular}
\begin{equation*}
Q^{i}=Q^{i}_{\chi}\cos\theta_{E_{6}}+Q^{i}_{\psi}\sin\theta_{E_{6}}.
\label{Chmixing}
\end{equation*}
\vspace{-1.5cm}
\caption{Charge assignments for the fields in several models.}
\label{charges}
\end{table}

Moreover, $v_{s}$ is now responsible for the spontaneous breaking of the $U(1)'$ symmetry, and hence the $\mu_{eff}$ term can be related to the breaking mechanism of a larger symmetry. The particle content of UMSSM is slightly richer than the MSSM. First of all, in addition to $S$ field, there should be also another gauge field associated with the $U(1)'$ group, which is denoted by $Z'$. Even though the current analyses provide strict bounds on $Z'$ ($M_{Z} \geq 2.7-3.3$ TeV \cite{Aaboud:2016cth}, $M_{Z}\geq 4.1$ TeV \cite{ATLAS:2017wce}), the signal processes in these analyses are based on the leptonic decay modes of $Z'$ as $Z'\rightarrow \bar{l}l$, where $l$ stands for the charged leptons of the first two families. However, as shown in a recent study \cite{Hicyilmaz:2017nzo}, $Z'$ may barely decay into two leptons; hence such strict bounds can be relaxed. Moreover, its neutral superpartner ($\tilde{B}'$) is also included in the low energy spectrum as required by SUSY. It is interesting that there is no specific mass bound on $\tilde{B}'$, and it can be as light as $\mathcal{O}(100)$ GeV consistent with the current experimental constraints \cite{Khalil:2015wua,DelleRose:2017ukx}. Since $\tilde{B}'$ is allowed to mix with the other neutralinos, the LSP neutralino may reveal its manifestation through the $U(1)'$ sector in the collider and DM direct detection experiments.

Since the MSSM fields are non-trivially charged under the extra $U(1)'$ group, the model may not be anomaly free. To avoid possible anomalies, one may include exotic fields whose participation into the triangle vertices leads to anomaly cancellation. However, such fields usually yield heavy exotic mass eigenstates at the low scale. If one chooses a superpotential in which these exotic fields do not interact with the MSSM fields directly, their effects in the sparticle spectrum are quite suppressed by their masses. In this case, the model effectively reduces to UMSSM without the exotic fields. After all, the superpotential is

\begin{equation}
W = Y_{u}\hat{Q}\hat{H}_{u}\hat{U}+Y_{d}\hat{Q}\hat{H}_{d}\hat{D}+Y_{e}\hat{L}\hat{H}_{d}\hat{E}+h_{s}\hat{S}\hat{H}_{d}\hat{H}_{u}
\label{suppot1}
\end{equation}
and the corresponding SSB Lagrangian can be written as

\begin{equation}
-\mathcal{L}^{\cancel{SUSY}}_{{\rm UMSSM}}=\mathcal{L}_{{\rm MSSM}}^{\cancel{SUSY}}+m_S^2|S|^2+M_{\tilde{B}'}\tilde{B}'\tilde{B}'+\left( A_{s}h_{s}SH_uH_d+h.c. \right)
\label{SSB_UMSSM}
\end{equation}

Employing Eqs.(\ref{suppot1} and \ref{SSB_UMSSM}), the Higgs potential can be obtained as

\begin{equation}
V^{{\rm tree}}=V_{F}^{{\rm tree}}+V_{D}^{{\rm tree}}+V_{\cancel{SUSY}}^{{\rm tree}}
\end{equation}
with

\begin{equation}
\setstretch{2.0}
\begin{array}{ll}
V_{F}^{{\rm tree}} & = | h_{s} |^{2} \left[| H_{u}H_{d}|^{2} + | S |^{2}\left( | H_{u}|^{2}+| H_{d}|^{2}  \right)   \right] \\
V_{D}^{{\rm tree}} & = \dfrac{g_{1}^{2}}{8}\left( | H_{u}|^{2}+| H_{d}|^{2}  \right)^{2}+\dfrac{g_{2}^{2}}{2}\left( |H_{u}|^{2}|H_{d}|^{2}-|H_{u}H_{d}|^{2}  \right) \\

& + \dfrac{g'^{2}}{2}\left( Q_{H_{u}}|H_{u}|^{2}+Q_{H_{d}}|H_{d}|^{2}+Q_{S}|S|^{2}  \right) \\

V_{\cancel{SUSY}}^{{\rm tree}} & = m^{2}_{H_{u}}|H_{u}|^{2}+m_{H_{d}}^{2}|H_{d}|^{2}+m_{S}^{2}|S|^{2}+\left(A_{s}h_{s}SH_{u}H_{d}+h.c. \right),
\end{array}
\end{equation}
which yields the following tree-level mass for the lightest CP-even Higgs boson mass:

\begin{equation}
m_h^2=M_Z^2\cos^22\beta+\left(v_u^2+v_d^2\right)\left[\frac{h_S^2\sin^22\beta}{2}+g_{Y^\prime}^2\left(Q_{H_{u}}\cos^2\beta+Q_{H_{d}}\sin^2\beta\right)\right].
\label{h0mass}
\end{equation}

The second term in the square paranthesis of Eq.(\ref{h0mass})
reflects the contribution from the $U(1)'$ sector, where $g_{Y'}$ is
the gauge coupling associated with $U(1)'$, $Q_{H_{u}}$ and
$Q_{H_{d}}$ are the charges of $H_{u}$ and $H_{d}$ under the $U(1)'$
group. After these contributions, the tree-level Higgs boson mass
can be obtained as large as about 140 GeV for low $\tan\beta$, while
it can be as heavy as about 115 GeV, when $\tan\beta$ is large
\cite{Hicyilmaz:2017nzo}. Similarly, other Higgs bosons receive
contributions from the $U(1)'$ sector, and the charged Higgs boson
mass can be obtained as tree-level as

\begin{equation}
m_{H^{\pm}}^{2}=M_{W}^{2}+\dfrac{\sqrt{2}h_{s}A_{s}v_{s}}{\sin(2\beta)}-\dfrac{1}{2}h_{s}^{2}(v_{d}^{2}+v_{u}^{2})
\end{equation}

In addition to the Higgs sector, the neutralino sector is also
enlarged in UMSSM. Since it has a field whose VEV breaks $U(1)'$
symmetry, its fermionic superpartner mixes with the MSSM
neutralinos, as in the NMSSM case. Moreover, since UMSSM possesses
an extra $U(1)$ symmetry, the gaugino partner of the gauge field
($Z'$) also mixes with the other neutralinos. In sum, there are six
neutralinos in UMSSM, and their masses and mixing can be given in
$(\tilde{B}',\tilde{B},\tilde{W},\tilde{h}_{u},\tilde{h}_{d},\tilde{S})$
basis as

\begin{equation}
\mathcal{M}_{\tilde{\chi}^{0}}=\left(\begin{array}{c:cccc:c}
M_{1}' & 0 & 0 & g'_{Y}Q_{H_{d}}v_{d} & g'_{Y}Q_{H_{u}}v_{u} & g'_{Y}Q_{S}v_{s} \\ \hdashline 0 & & & & & 0 \\ 0 & &&& & 0 \\ g'_{Y}Q_{H_{d}}v_{d} & \multicolumn{4}{c:}{\mathcal{M}_{\tilde{\chi}^{0}}^{{\rm MSSM}}(\mu=\mu_{eff})}& -\dfrac{1}{\sqrt{2}}h_{s}v_{u} \\ g'_{Y}Q_{H_{u}}v_{u} &&&&& -\dfrac{1}{\sqrt{2}}h_{s}v_{d} \\ \hdashline &&&&&\\ g'_{Y}Q_{S}v_{s} &    0 & 0 & -\dfrac{1}{\sqrt{2}}h_{s}v_{u} & -\dfrac{1}{\sqrt{2}}h_{s}v_{d} & 0
\end{array}
\right)
\end{equation}
where $M'_{1}$ is the SSB mass of $\tilde{B}'$, and the first row
and column code the mixing of $\tilde{B}'$ with the other
neutralinos. The middle part represents the MSSM neutralino masses
and mixing, while the last column and row displays the mass and
mixing for the MSSM singlet field as in the case of NMSSM. Similar
to NMSSM, UMSSM does not propose any new charged particle; hence,
the chargino sector remains the same as that in MSSM.

\section{Scanning Procedure and Experimental Constraints}
\label{sec:scan}

We have employed SPheno 3.3.3 package \cite{Porod:2003um} obtained with SARAH 4.5.8 \cite{Staub:2008uz}. In this package, the weak scale values of the gauge and Yukawa couplings are evolved to the unification scale $M_{{\rm GUT}}$ via the renormalization group equations (RGEs). $M_{{\rm GUT}}$ is determined by the requirement of the gauge coupling unification, described as $g_{1}=g_{2}$ for CMSSM and NMSSM, while it is as $g_{1}=g_{2}=g_{Y'}$ for UMSSM. {Note that the UMSSM framework is not anomaly-free, but its RGE's are being used, since the U(1)' models reduce to UMSSM effectively due to possible heavy exotic states. This treatment can be improved with the inclusion of such exotic states in the RGEs.} Even though $g_{3}$ does not appear in these conditions for $M_{{\rm GUT}}$, it needs to take part in the gauge coupling unification condition. Concerning the contributions from the threshold corrections to the gauge couplings at the GUT scale arising from some unknown breaking mechanisms of the GUT gauge group, $g_{3}$ receives the largest contributions \cite{Hisano:1992jj}, and it is allowed to deviate from the unification up to about $3\%$. If a solution does not require this condition within this allowance, SPheno does not generate an output for such solutions by default. Hence, the existence of an output file guarantees that the solutions {are} compatible with the unification condition, and $g_{3}$ deviates no more than $3\%$. With the boundary conditions given at $M_{{\rm GUT}}$, all the SSB parameters along with the gauge and Yukawa couplings are evolved back to the weak scale. Note that each model yields different RGEs coded by SARAH in different model packages for SPheno. We employ the packages called after the model names as MSSM, NMSSM and UMSSM. During our numerical investigation, we have performed random scans over the following parameter spaces of CMSSM, NMSSM and UMSSM:

\begin{equation}
\centering
\setstretch{1.5}
\scalebox{0.80}{$
\begin{array}{ccc|ccc|ccc}
\multicolumn{3}{c|}{{\rm CMSSM}} & \multicolumn{3}{c|}{{\rm NMSSM}} & \multicolumn{3}{c}{{\rm UMSSM}} \\ \hline
0 \leq & m_{0} & \leq 5~{\rm (TeV)} & 0 \leq & m_{0} & \leq 3~{\rm (TeV)} & 0 \leq & m_{0} & \leq 3~{\rm (TeV)} \\
0 \leq & M_{1/2} & \leq 5~{\rm (TeV)} &0 \leq & M_{1/2} & \leq 3~{\rm (TeV)} & 0 \leq & M_{1/2} & \leq 3~{\rm (TeV)} \\
1.2 \leq & \tan\beta & \leq 50 & 1.2 \leq & \tan\beta & \leq 50 & 1.2 \leq & \tan\beta & \leq 50 \\
-3 \leq & A_{0}/m_{0} & \leq 3 & -3 \leq & A_{0}/m_{0} & \leq 3 & -3 \leq & A_{0}/m_{0} & \leq 3 \\
& \mu > 0 & & 0 \leq & h_{s} & \leq 0.7 & 0 \leq & h_{s} & \leq 0.7 \\
& & & 1 \leq & v_{s} & \leq 25~{\rm (TeV)} & 1 \leq & v_{s} & \leq 25~{\rm (TeV)} \\
&&& -10 \leq & A_{s},A_{\kappa} & \leq 10~{\rm (TeV)} & -10 \leq & A_{s} & \leq 10~{\rm (TeV)} \\
&&& 0 \leq & \kappa & \leq 0.7 & -\dfrac{\pi}{2} \leq & \theta_{E_{6}} & \leq \dfrac{\pi}{2}
\end{array}$}
\label{paramSP}
\end{equation}
where $m_{0}$ is the universal spontaneous supersymmetry breaking (SSB) mass term for the matter scalars. This mass term is also set as $m_{H_{u}}=m_{H_{d}}=m_{0}$ in CMSSM, while $m_{H_{u}}$ and $m_{H_{d}}$ are calculated through the EWSB conditions, which leads to $m_{H_{u}}\neq m_{H_{d}}\neq m_{0}$ in NMSSM and UMSSM. Similarly, $M_{1/2}$ is the universal SSB mass term for
the gaugino fields, which includes one associated with the $U(1)'$
gauge group in UMSSM. $\tan\beta=\langle v_{u} \rangle / \langle
v_{d} \rangle$ is the ratio of VEVs of the MSSM Higgs doublets,
$A_{0}$ is the SSB term for the trilinear scalar interactions
between the matter scalars and MSSM Higgs fields. Similarly, $A_{s}$
is the SSB interaction between the $S$ and $H_{u,d}$ fields, and
$A_{\kappa}$ is the SSB term for the triple self interactions of the
$S$ fields. $h_{s}$ and $\kappa$ have defined before. Note that
$\kappa =0$ in the UMSSM case. Finally, $v_{s}$ denotes the VEV of
$S$ fields. Recall that the $\mu-$term of MSSM is dynamically
generated such that $\mu = h_{s}v_{s}/\sqrt{2}$. Its sign is
assigned as a free parameter in MSSM, since REWSB condition can
determine its value but not sign. For simplicity, we forced it  to
be positive in NMSSM and in UMSSM by $h_{s}$ and $v_{s}$. Finally,
we set the top quark mass to its central value ($m_{t} = 173.3$ GeV)
\cite{Group:2009ad}. Note that the sparticle spectrum is not {very}
sensitive in one or two sigma variation in the top quark mass
\cite{Gogoladze:2011db}, but it can shift the Higgs boson mass by
$1-2$ GeV  \cite{Gogoladze:2011aa}.

{The requirement of radiative electroweak symmetry breaking (REWSB) provides important theoretical constraints, since it excludes the solutions with $m_{H_{u}}^{2}=m_{H_{d}}^{2}$ \cite{Martin:1997ns}. Besides, based on our previous experience from the numerical analyses over the parameter spaces of various SUSY models, REWSB requires $m_{H_{u}}^{2}$ to be more negative than $m_{H_{d}}^{2}$ at the low scale (see for instance \cite{Gogoladze:2011aa}). In addition, the solutions are required to bring consistent values for the gauge and Yukawa couplings, gauge boson masses, top quark mass \cite{Ibanez:1982fr} etc. Such constraints are being imposed into SPheno by default. In addition, the solutions must not yield color and/or charge breaking minima, which restricts the trilinear scalar interaction coupling in our scans as $|A_{0}| \leq 3m_{0}$, where $m_{0}$ is the universal mass term at the GUT scale for the SUSY scalars.}  Another important constraint comes from the relic abundance of the stable charged particles \cite{Nakamura:2010zzi},
which excludes the regions where charged SUSY particles such as stau and stop become the lightest supersymmetric particle (LSP). In our scans, we allow only the solutions for which one of the neutralinos is the LSP and REWSB condition is satisfied.

In scanning the parameter space, we use our interface, which employs
Metropolis-Hasting algorithm described in \cite{Belanger:2009ti}.
After collecting the data, we impose the mass bounds on all the
sparticles \cite{Agashe:2014kda}, and the constraint from the rare
B-decays such as $B_{s}\rightarrow \mu^{+}\mu^{-}$
\cite{Aaij:2012nna}, $B_{s}\rightarrow X_{s}\gamma$
\cite{Amhis:2012bh}, and $B_{u}\rightarrow \tau \nu_{\tau}$
\cite{Asner:2010qj}. In addition,  the WMAP bound
\cite{Hinshaw:2012aka} on the relic abundance of neutralino LSP
within $5\sigma$ uncertainty. Note that the current results from the
Planck satellite \cite{Ade:2015xua} allow more or less a similar
range for the DM relic abundance within $5\sigma$ uncertainty, when
one takes the uncertainties in calculation. These experimental
constraints can be summarized as follows:

\begin{equation}
\setstretch{1.8}
\begin{array}{l}
m_h  = 123-127~{\rm GeV}
\\
m_{\tilde{g}} \geq 1.8~{\rm TeV}
\\
M_{Z'} \geq 2.5 ~{\rm TeV} \\
0.8\times 10^{-9} \leq{\rm BR}(B_s \rightarrow \mu^+ \mu^-)
  \leq 6.2 \times10^{-9} \;(2\sigma)
\\
m_{\tilde{\chi}_{1}^{0}} \geq 103.5~{\rm GeV} \\
m_{\tilde{\tau}} \geq 105~{\rm GeV} \\
2.99 \times 10^{-4} \leq
  {\rm BR}(B \rightarrow X_{s} \gamma)
  \leq 3.87 \times 10^{-4} \; (2\sigma)
\\
0.15 \leq \dfrac{
 {\rm BR}(B_u\rightarrow\tau \nu_{\tau})_{\rm MSSM}}
 {{\rm BR}(B_u\rightarrow \tau \nu_{\tau})_{\rm SM}}
        \leq 2.41 \; (3\sigma) \\
   0.0913 \leq \Omega_{{\rm CDM}}h^{2} \leq 0.1363~(5\sigma)
\label{constraints}
\end{array}
\end{equation}

{In addition to those listed above, we also employ the constraints on the SM-like Higgs boson decay processes obtained from the ATLAS \cite{ATLAS:2014aga} and CMS \cite{Chatrchyan:2013iaa} analyses. We expect a strong impact from current bounds for ${\rm BR}(B\rightarrow X_{s}\gamma)$ on the parameter space. Figure \ref{fig:consonHpm} displays the results for the impacts from the $B\rightarrow X_{s}\gamma$ and $h\rightarrow ZZ$ processes (where $h$ denotes the SM-like Higgs boson) with plots in the ${\rm BR}(B_{s}\rightarrow X_{s}\gamma)-m_{H^{\pm}}$ and ${\rm BR}(h\rightarrow ZZ)-m_{H^{\pm}}$ planes for CMSSM. All points are compatible with the REWSB and neutralino LSP conditions. Green points satisfy the mass bounds and the constraints from the rare decays of B-meson. The blue points form a subset of green and they are consistent with the constraints from the SM-like Higgs boson decay processes. {Note that the constraint from ${\rm BR}(B_{s}\rightarrow X_{s}\gamma)$ is not applied in the left plane, but the bounds from this processes are represented with the horizontal solid lines. Similarly, the constraint from ${\rm BR}(h\rightarrow ZZ)$ is not employed in the right plane, and the horizontal lines indicate the experimental bounds {($0.024 \leq {\rm BR}(h\rightarrow ZZ) \leq 0.029$) { within $2\sigma$ uncertainty\cite{Amhis:2012bh,ATLAS:2014aga,Chatrchyan:2013iaa}.}} As we expect,  the constraint from the ${\rm BR}(B_{s}\rightarrow X_{s}\gamma)$ process excludes a significant portion of the parameter space (green points below the bottom horizontal line); however, its impact barely affects the mass bound on the charged Higgs boson. We can find the $m_{H^{\pm}} \gtrsim 800$ GeV allowed by this constraints (see also \cite{Domingo:2007dx}). On the other hand, the main impact on the parameter space (and hence on the charged Higgs boson mass) comes from those for the SM-like Higgs boson decays. The ${\rm BR}(h\rightarrow ZZ)-m_{H^{\pm}}$ plane shows that the $h\rightarrow ZZ$ process excludes more than half of the parameter space (green and blue points below the bottom horizontal line in the right panel). According to the results in the ${\rm BR}(h\rightarrow ZZ)-m_{H^{\pm}}$ plane, the solutions with $m_{H^{\pm}} \lesssim 2$ TeV are excluded by the $h\rightarrow ZZ$ process. Note that the universal boundary conditions of CMSSM restricts results more, and if non-universality is employed, the lower mass for the charged Higgs boson can be found at about 1 TeV consistently with the current constraints including that from the $h\rightarrow ZZ$ process \cite{Li:2016ucz}. Thus, our results will also mean that some possible signal channels require to impose non-universal boundary conditions in the MSSM framework to be available.}

\begin{figure}[h!]
\centering
\subfigure{\includegraphics[scale=0.4]{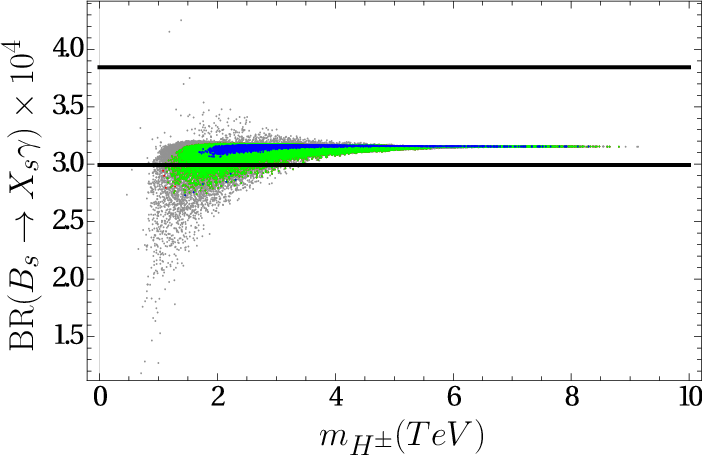}}
\subfigure{\includegraphics[scale=0.4]{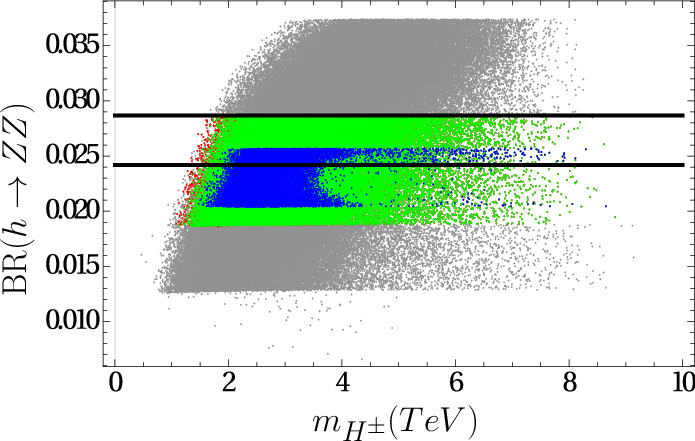}}
\caption{CMSSM plots in the ${\rm BR}(B_{s}\rightarrow X_{s}\gamma)-m_{H^{\pm}}$ and ${\rm BR}(h\rightarrow ZZ)-m_{H^{\pm}}$ planes. All points are compatible with the REWSB and neutralino LSP conditions. Green points satisfy the mass bounds and the constraints from the rare decays of B-meson. The blue points form a subset of green and they are consistent with the constraints from the SM-like Higgs boson decay processes. {Note that the constraint from ${\rm BR}(B_{s}\rightarrow X_{s}\gamma)$ is not applied in the left plane, but the bounds from this processes are represented with the horizontal solid lines. Similarly, the constraint from ${\rm BR}(h\rightarrow ZZ)$ is not employed in the right plane, and the horizontal lines indicate the experimental bounds from this process ($0.024 \leq {\rm BR}(h\rightarrow ZZ) \leq 0.029$) { within $2\sigma$ uncertainty\cite{Amhis:2012bh,ATLAS:2014aga,Chatrchyan:2013iaa}.}}}
\label{fig:consonHpm}
\end{figure}

\begin{figure}[hb!]
\centering
\subfigure{\includegraphics[scale=0.4]{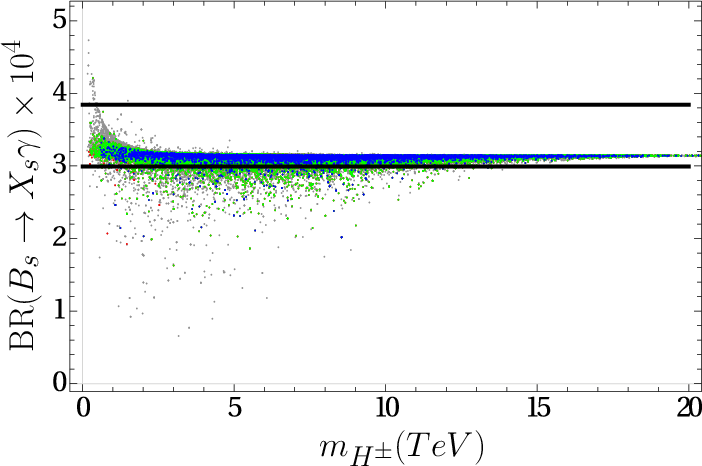}}
\subfigure{\includegraphics[scale=0.4]{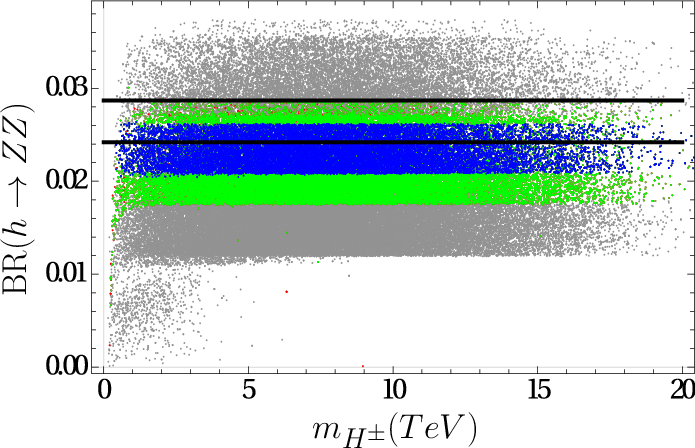}}
\subfigure{\includegraphics[scale=0.4]{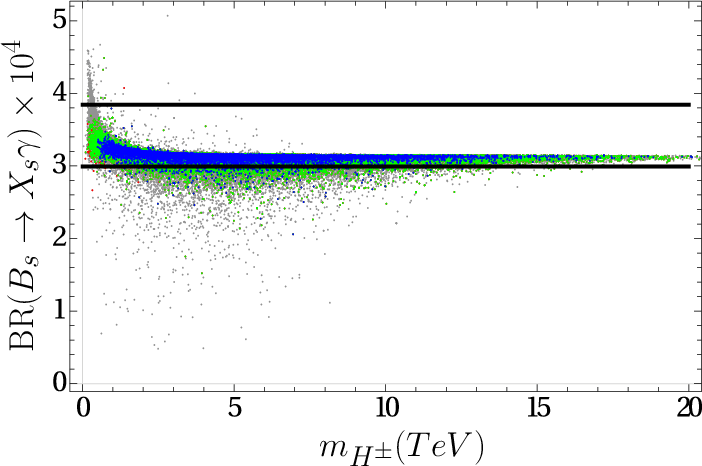}}
\subfigure{\includegraphics[scale=0.4]{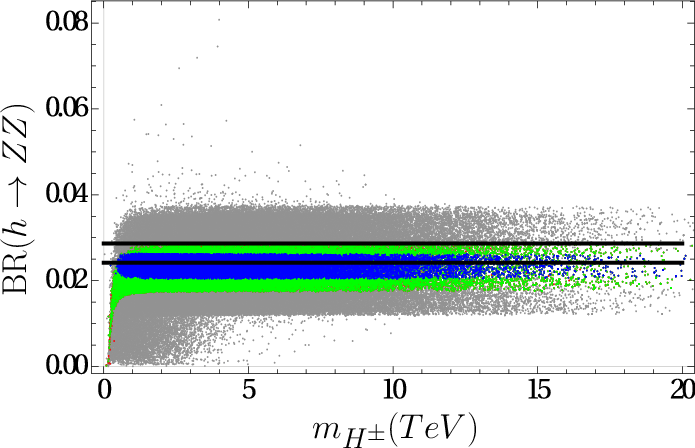}}
\caption{Plots for the impacts of ${\rm BR}(B_{s}\rightarrow X_{s}\gamma)$ (left) and ${\rm BR}(h\rightarrow ZZ)$ (right) on the parameter space of NMSSM (top) and UMSSM (bottom). The color coding is the same as Figure \ref{fig:consonHpm}.}
\label{fig:NMSSMUMSSM}
\end{figure}

{On the other hand, one can investigate the impacts of these constraints on the fundamental parameter space of NMSSM and UMSSM as shown in Figure \ref{fig:NMSSMUMSSM} with plots for the impacts of ${\rm BR}(B_{s}\rightarrow X_{s}\gamma)$ (left) and ${\rm BR}(h\rightarrow ZZ)$ (right) on the parameter space of NMSSM (top) and UMSSM (bottom). The color coding is the same as Figure \ref{fig:consonHpm}. The results in Figure \ref{fig:NMSSMUMSSM} shows that the constraint from $h\rightarrow ZZ$ has a strong impact in both models. However, even though this constraint excludes more than half of the solutions, it does not bound the charged Higgs mass from below, in contrast to CMSSM.} {Note that lighter mass scales for the Charged Higgs boson can be allowed, if one employs relatively milder bounds from the $h\rightarrow ZZ$ process.}

We have emphasized the bounds on the Higgs boson \cite{:2012gk} and
the gluino \cite{gluinoATLAS}, since they have drastically changed
since the LEP era. {We have employed the two-loop RGEs in calculation of the Higgs boson mass. The uncertainty in the Higgs boson mass calculation arises mostly from the uncertainties in values of the strong gauge coupling and top quark masses, which yield about 3 GeV deviation in the Higgs boson mass calculation \cite{Degrassi:2002fi}. {In addition, the large SUSY scale ($M_{{\rm SUSY}}$) worsens the uncertainty in the Higgs boson mass calculation \cite{Allanach:2018fif}.} Note that there are more precise calculations available to improve the results for the Higgs boson mass (see for instance \cite{Staub:2017jnp}).} In addition, we have employed the LEP II mass bounds on the lightest chargino and stau. Although the mass bounds have recently been updated on these particles \cite{CMS:2017fdz,Sirunyan:2017zss}, these bounds are model dependent and based on specific decay channels of the chargino. While we employ the LEP II bounds on our plots, their masses will be discussed briefly later. In addition, the current mass bounds on $Z'$ is established as $M_{Z'} \gtrsim 4.1$ TeV \cite{Aaboud:2016cth}. On the other hand, this bound can vary model dependently, and a most recent study \cite{Araz:2017wbp} has shown that $M_{Z'} \gtrsim 2.5$ TeV can survive, if $Z'$ is leptophobic. Based on our previous study \cite{Hicyilmaz:2017nzo}, which was conducted in the similar parameter space, the leptonic decay processes of $Z'$ is found as ${\rm BR}(Z'\rightarrow ll)\lesssim 14\%$. {Since the leptonic decays of $Z'$ are found rather low, we set the mass bound on $Z'$ as $M_{Z'} \geq 2.5$ TeV. {The mass bound on $Z'$ depends on the gauge coupling associated with $U(1)'$ group which varies with $\theta_{E_{6}}$. Thus, some of the solutions represented in our study can be excluded by the experimental analyses \cite{Aaboud:2016cth}.}}


When the LSP is required to be one of the neutralinos, the DM relic abundance constraint will be highly effective to shape the fundamental parameter space, since the relic abundance of LSP neutralino is usually realized greater than the current measurement over most of the fundamental parameter space. Once one can identify the regions compatible with the current WMAP and Planck results, they can be analyzed further against the results from the direct detection \cite{Akerib:2016lao}, indirect detection \cite{Atwood:2009ez} and collider experiments \cite{Buchmueller:2017qhf}. However, all these detailed analyses are out of the scope of our study. We apply only the relic abundance constraint on the LSP neutralino to show that the regions of interest for the charged Higgs boson phenomenology can also be compatible with the relic abundance bound from the current measurements, and they can be also tested under the light of the DM constraints in possible future studies. In this context, the DM implications obtained within our analyses can be improved with more thorough analyses devoted to DM. The DM observables in our scan are calculated by micrOMEGAs \cite{Belanger:2006is} obtained by SARAH \cite{Staub:2008uz}.

We will apply the constraints mentioned in this section subsequently, and thus, before concluding this section, it might be necessary to mention the color convention which we will use in the next sections in presenting the results. The following is the list that summarizes what color satisfies which constraints:

\begin{itemize}

\item[Grey:] REWSB and Neutralino LSP conditions.

\item[Red:] REWSB, neutralino LSP and Higgs boson mass constraint.

\item[Green:] REWSB, neutralino LSP, Higgs boson mass constraint, SUSY particle mass bounds, and B-physics constraints.

\item[Blue:] REWSB, neutralino LSP, Higgs boson mass constraint, SUSY particle mass bounds, B-physics constraints, LHC constraints on the Higgs boson couplings.

\item[Black:] REWSB, neutralino LSP, Higgs boson mass constraint, SUSY particle mass bounds, B-physics constraints, LHC constraints on the Higgs boson couplings, and WMAP and Planck constraints on the relic abundance of neutralino LSP within $5\sigma$.
\end{itemize}

{From black to grey, each color is always on top of the previous one in the order as listed above in a way that the black always stays on top of all other colors in the plots.}

\section{Fundamental Parameter Space and Mass Spectrum}
\label{sec:spectrum}

\begin{figure}[h!]
\centering
\includegraphics[scale=0.4]{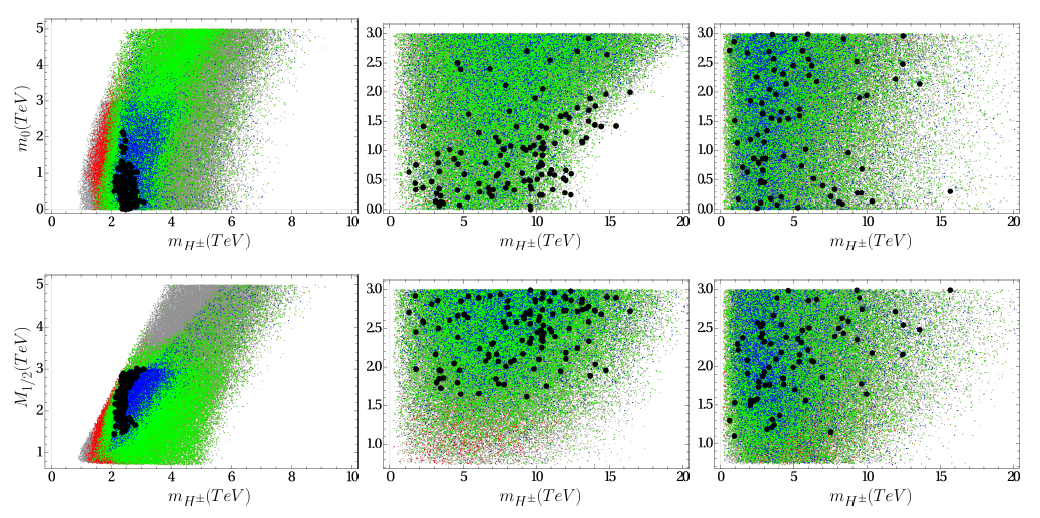}
\caption{Plots in the $m_{0}-m_{H^{\pm}}$ and $M_{1/2}-m_{H^{\pm}}$ planes for CMSSM (left panel), NMSSM (middle panel) and UMSSM (right panel). All points are consistent with the REWSB and neutralino LSP. Our color convention is as listed at the end of Section \ref{sec:scan}.}
\label{fig:m0m21_Hpm}
\end{figure}

In this section, we consider the fundamental parameter space, shaped by the experimental constraints discussed in the previous section, and discuss the charged Higgs mass along with the mass spectrum for other particles, which might be relevant to decay modes {of} the charged Higgs boson. Figure \ref{fig:m0m21_Hpm} displays our results with plots in the $m_{0}-m_{H^{\pm}}$ and $M_{1/2}-m_{H^{\pm}}$ planes for CMSSM (left panel), NMSSM (middle panel) and UMSSM (right panel). All points are consistent with the REWSB and neutralino LSP. Our color convention is as listed at the end of Section \ref{sec:scan}. In the CMSSM case, as seen from the left panel, the charged Higgs can be as heavy as about 8 TeV in the range of the fundamental parameters given in Eq.(\ref{paramSP}).

These results in CMSSM arise from the fact that CMSSM yields mostly bino-like LSP neutralino $\mu \gg M_{2}\sim 2M_{1}$ \cite{Haber:1984rc}, where $\mu$ is the Higgsino mass parameter, while $M_{2}$ and $M_{1}$ are the SSB masses of Wino and Bino, respectively. The problem with bino-like DM is that its relic abundance is usually much larger than the current measurements of the WMAP \cite{Hinshaw:2012aka} and Planck \cite{Ade:2015xua} satellites, and one needs to identify some coannihilation channels which take part in reducing LSP neutralino's relic abundance down to the current ranges \cite{Gogoladze:2009bn}. However, the void direct signals for supersymmetry from the LHC experiments yield quite heavy spectrum in the low scale implications of CMSSM, and the mass scales are usually out of the possible coannihilation scenarios. On the other hand, even if the neutralino sector is extended only a single flavor, the region of the parameter space allowed by the DM observations becomes quite wide open \cite{DelleRose:2017ukx}, as a result of mixing the extra flavor state with the MSSM neutralinos. However, as seen from the bottom left panel of Figure \ref{fig:m0m21_Hpm}, the charged Higgs boson cannot be lighter than about 1.5 TeV when one employs the LHC constraints (green). {Even though we do not find solutions for $m_{H^{\pm}}\lesssim 2$ TeV after applying all the constraints listed in Section \ref{sec:scan}, some recent studies \cite{Han:2016gvr} have shown that $m_{H^{\pm}}\gtrsim 1.5$ TeV can be consistent with the DM constraints.} Such a lower bound on the charged Higgs boson mass mostly arises from the rare B-meson decay process, $B_{s}\rightarrow X_{s}\gamma$, where $X_{s}$ is a suitable bound state of the strange quark. The strong agreement between the experimental measurements (${\rm BR}^{{\rm exp}}(B_{s}\rightarrow X_{s}\gamma)=(3.43\pm 0.22)\times 10^{-4}$ \cite{Amhis:2012bh}) and the Standard Model prediction (${\rm BR}^{{\rm SM}}(B_{s}\rightarrow X_{s}\gamma)=(3.15\pm 0.23)\times 10^{-4}$ \cite{Misiak:2006zs}) strongly {enforces} a lower bound on the charged Higgs boson mass. {However; as have been shown before, these constraints from the rare B-meson decays restrict the charged Higgs boson mass as $m_{H^{\pm}} \gtrsim 800$ GeV. The strongest restriction comes from the $h\rightarrow ZZ$ process, which excludes the solutions with $m_{H^{\pm}} \lesssim 2$ TeV (blue points). In contrast to the results in CMSSM, the mass range of the charged Higgs boson is quite wide in NMSSM and UMSSM, as is seen from the middle and right panels respectively, and the solutions can yield $m_{H^{\pm}}$ from about 1 to 15 TeV, after the experimental constraints are employed. This wide mass range partly arises from the non-universality in $m_{H_{d}}$ and $m_{H_{u}}$ in NMSSM and UMSSM.} In addition, such a wide region allowed by the DM observations reflects the significant effect of extending the neutralino sector of MSSM even with one extra flavor state. The recent studies have shown that new flavor states, which are allowed to mix with the MSSM neutralinos can significantly alter the DM implications \cite{DelleRose:2017ukx}. The fundamental parameter space for NMSSM and UMSSM are restricted based on our previous studies \cite{Hicyilmaz:2017nzo,Hicyilmaz:2016kty}, which explored the regions with acceptable fine-tuning in the fundamental parameter space of UMSSM.


\begin{figure}[h!]
\centering
\includegraphics[scale=0.4]{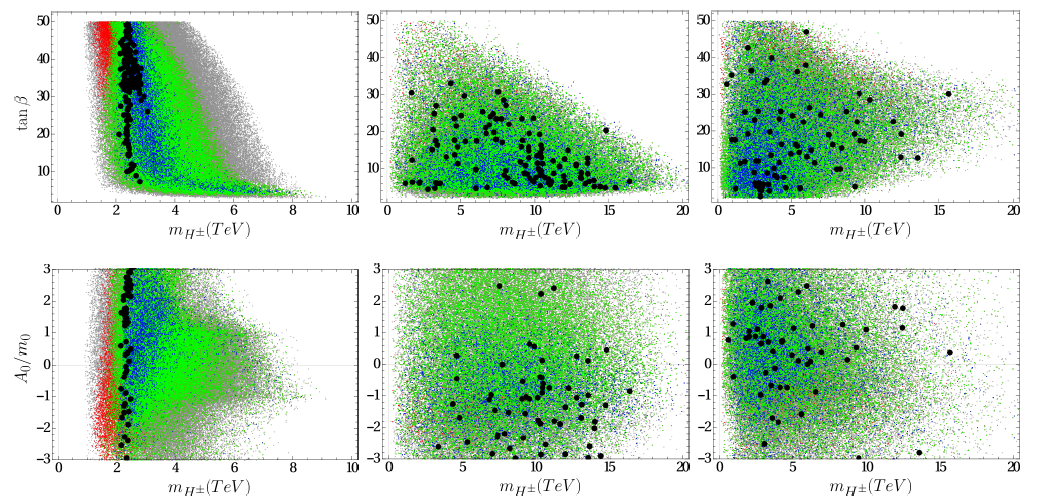}
\caption{Plots in the $\tan\beta-m_{H^{\pm}}$ and $A_{0}/m_{0}-m_{H^{\pm}}$ planes for CMSSM (left panel), NMSSM (middle panel) and UMSSM (right panel). The color coding is the same as Figure \ref{fig:m0m21_Hpm}.}
\label{fig:tanA0_Hpm}
\end{figure}

The other fundamental parameters are $A_{0}$ and $\tan\beta$, and
the results in terms of these parameters are represented in Figure
\ref{fig:tanA0_Hpm} with plots in the $\tan\beta-m_{H^{\pm}}$ and
$A_{0}/m_{0}-m_{H^{\pm}}$ planes for CMSSM (left panel), NMSSM
(middle panel) and UMSSM (right panel). The color coding is the same
as Figure \ref{fig:m0m21_Hpm}. As seen from the top panels,
$\tan\beta$ is a strong parameter in $m_{H^{\pm}}$, and one can
realize heavy charged Higgs boson only when $\tan\beta \lesssim 10$
in the CMSSM and NMSSM frames, while it is restricted to moderate
values as $20 \lesssim \tan\beta \lesssim 30$ in UMSSM. On the other
hand, there is no specific restriction in $A_{0}$, and as seen from
the bottom panels, it is possible to realize whole allowed range of
$m_{H^{\pm}}$ for any $A_{0}$.

\begin{figure}[h!]
\centering
\includegraphics[scale=0.4]{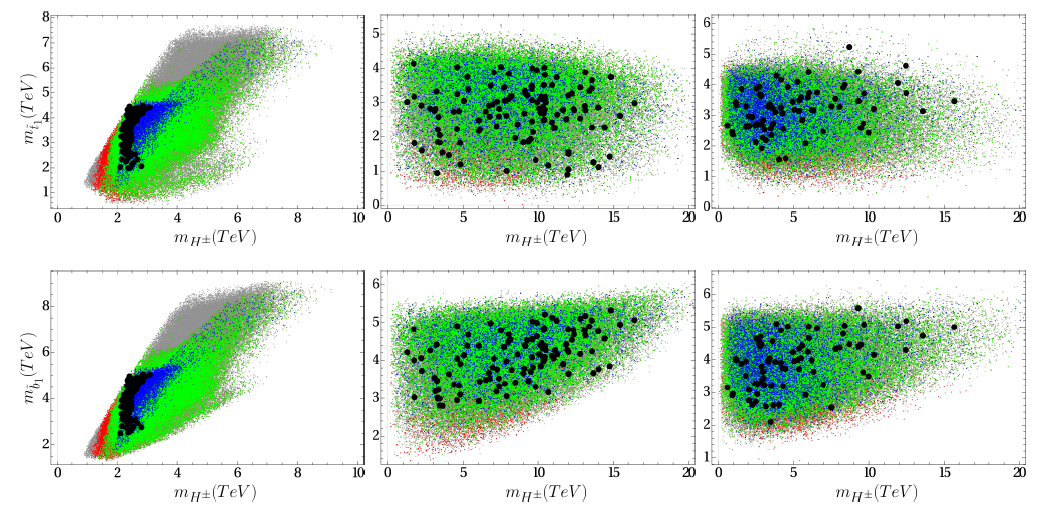}
\caption{Plots in the $m_{\tilde{t}_{1}}-m_{H^{\pm}}$ and $m_{\tilde{b}_{1}}-m_{H^{\pm}}$ planes for CMSSM (left panel), NMSSM (middle panel) and UMSSM (right panel). The color coding is the same as Figure \ref{fig:m0m21_Hpm}.}
\label{fig:stopsbottom_Hpm}
\end{figure}

After presenting the fundamental parameter space of the models,
we consider the mass spectrum, {that} reveals {which} particles the
charged Higgs boson may kinematically be allowed to decay. First we
present the stop and sbottom masses in Figure
\ref{fig:stopsbottom_Hpm} with plots in the
$m_{\tilde{t}_{1}}-m_{H^{\pm}}$ and $m_{\tilde{b}_{1}}-m_{H^{\pm}}$
planes for CMSSM (left panel), NMSSM (middle panel) and UMSSM (right
panel). If the solutions with $m_{H^{\pm}}\gtrsim
m_{\tilde{t}}+m_{\tilde{b}}$ can be realized, then the charged Higgs
boson can participate in the processes $H^{\pm}\rightarrow
\tilde{t}\tilde{b}$. As seen from the middle and right
panels, {if the charged Higgs boson is allowed to be heavy enough, there is a possibility for the decay process} $H^{\pm}\rightarrow
\tilde{t}\tilde{b}$. However, since the relevant background
generated by the decay processes involving with the top-quark
significantly suppresses the possible signals from stop
\cite{Cici:2016oqr}, such decays of the charged Higgs boson may not
provide a detectable track.

\begin{figure}[hb!]
\centering
\includegraphics[scale=0.4]{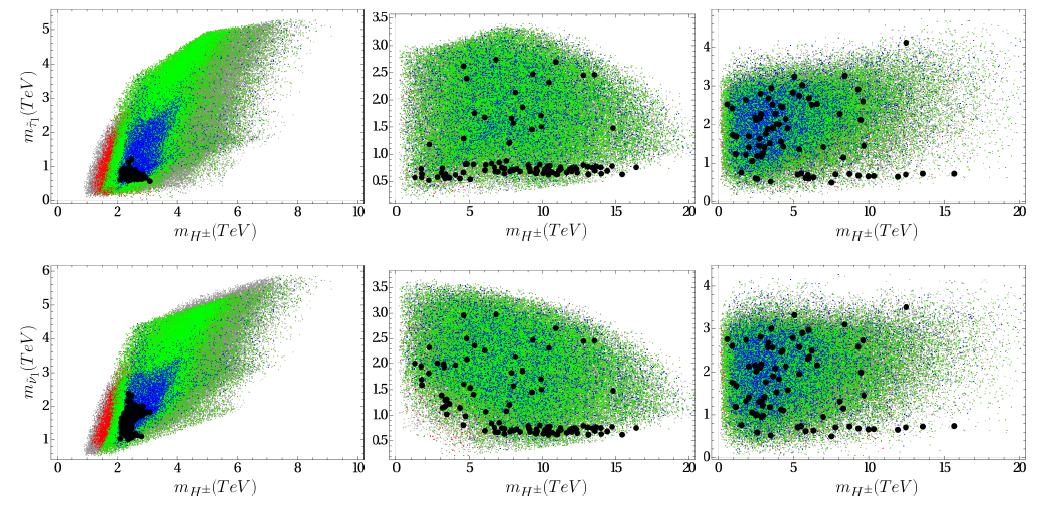}
\caption{Plots in the $m_{\tilde{\tau}_{1}}-m_{H^{\pm}}$ and $m_{\tilde{\nu}}-m_{H^{\pm}}$ planes for CMSSM (left panel), NMSSM (middle panel) and UMSSM (right panel). The color coding is the same as Figure \ref{fig:m0m21_Hpm}.}
\label{fig:stausneutrino_Hpm}
\end{figure}

Figure \ref{fig:stausneutrino_Hpm} displays our results with another
pair of supersymmetric particles, stau and sneutrino, which the
charged Higgs boson can decay, with plots in the
$m_{\tilde{\tau}_{1}}-m_{H^{\pm}}$ and $m_{\tilde{\nu}}-m_{H^{\pm}}$
planes for CMSSM (left panel), NMSSM (middle panel) and UMSSM (right
panel). The color coding is the same as Figure \ref{fig:m0m21_Hpm}.
The current bound on a charged slepton can be expressed as
$m_{\tilde{\tau}}\gtrsim 400$ GeV \cite{Calibbi:2014pza}. Such
bounds rely on the chargino-neutralino production, which differs
from model to model; thus it can vary depending on the mass
spectrum. Considering the model dependence of such bounds, even if
we employed the LEP2 bounds, the LHC and DM constraints bound the
stau mass from below as $m_{\tilde{\tau}}\gtrsim 500$ GeV in all
CMSSM, NMSSM and UMSSM as seen from the top panels. Similarly the
sneutrino mass is also bounded as $m_{\tilde{\nu}}\gtrsim 1$ TeV.

\begin{figure}[h!]
\centering
\includegraphics[scale=0.4]{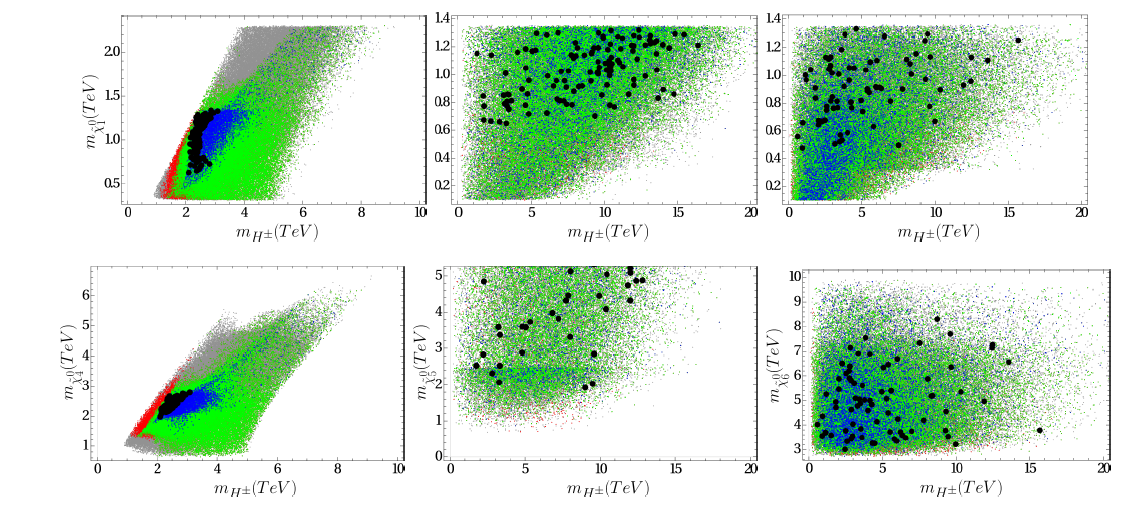}
\caption{Plots in the $m_{\tilde{\chi}_{1}^{0}}-m_{H^{\pm}}$ and $m_{\tilde{\chi}_{i}^{0}}-m_{H^{\pm}}$ planes for CMSSM (left panel), NMSSM (middle panel) and UMSSM (right panel), where $i$ stands for the number identifying the heaviest neutralino in models as $i=4$ for CMSSM, $i=5$ for NMSSM, and $i=6$ for UMSSM. The color coding is the same as Figure \ref{fig:m0m21_Hpm}.}
\label{fig:neutralinos_Hpm}
\end{figure}

Figure \ref{fig:neutralinos_Hpm} shows the neutralino masses and the
charged Higgs boson mass with plots in the
$m_{\tilde{\chi}_{1}^{0}}-m_{H^{\pm}}$ and
$m_{\tilde{\chi}_{i}^{0}}-m_{H^{\pm}}$ planes for CMSSM (left panel),
NMSSM (middle panel) and UMSSM (right panel), where $i$ stands for
the number identifying the heaviest neutralino in models as $i=4$
for CMSSM, $i=5$ for NMSSM, and $i=6$ for UMSSM. The color coding is
the same as Figure \ref{fig:m0m21_Hpm}. All models allow the LSP neutralino to be only as heavy as about 1.5 TeV. The upper bound on the neutralino LSP mass arises from the range assigned to $M_{1/2}$ in scanning the fundamental parameter spaces of the models. The lower bound, on the other hand, arises mostly from the heavy bound on the gluino mass, while the other constraints may also have minor effects. Without the relic density constraint, the neutralino LSP mass can be as low as about 400 GeV in CMSSM, while lower bound can be as low as about 100 GeV in NMSSM and UMSSM (blue). However, the WMAP and Planck bounds on the relic abundance of neutralino LSP can be satisfied when $m_{\tilde{\chi}_{1}^{0}} \gtrsim 500$ GeV in all models. The heaviest neutralino $\tilde{\chi}_{i}^{0}$ in CMSSM ($i=4$) cannot be lighter than about 2 TeV, while its mass is bounded from above as $m_{\tilde{\chi}_{4}^{0}}\lesssim 3$ TeV. While NMSSM and UMSSM reveal similar bound from below at about 500 GeV, the heaviest neutralino mass in these models can be realized in multi-TeV scale as $m_{\tilde{\chi}_{5}^{0}} \lesssim 5$ TeV in NMSSM and $m_{\tilde{\chi}_{5}^{0}} \lesssim 10$ TeV in UMSSM.

\begin{figure}[h!]
\centering
\includegraphics[scale=0.4]{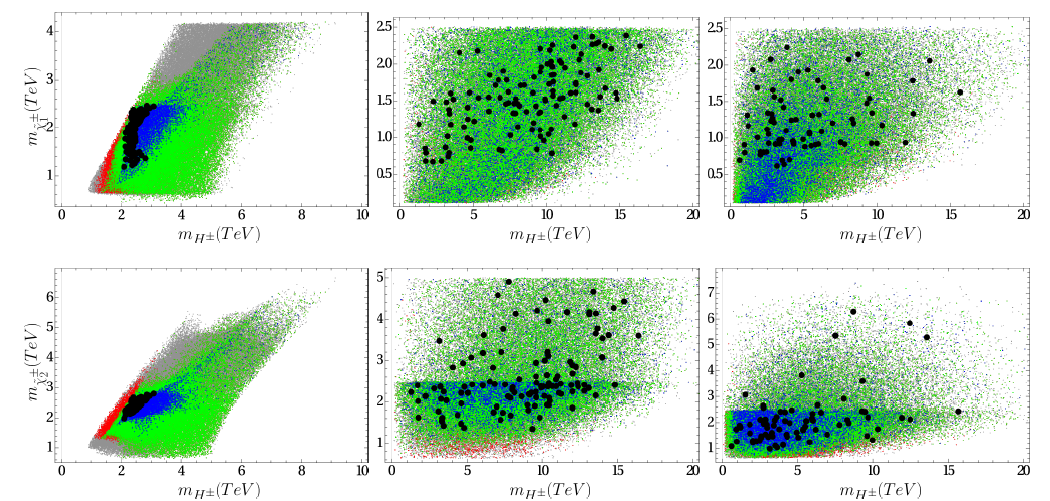}
\caption{Plots in the $m_{\tilde{\chi}_{1}^{\pm}}-m_{H^{\pm}}$ and $m_{\tilde{\chi}_{2}^{\pm}}-m_{H^{\pm}}$ planes for CMSSM (left panel), NMSSM (middle panel) and UMSSM (right panel). The color coding is the same as Figure \ref{fig:m0m21_Hpm}.}
\label{fig:charginos_Hpm}
\end{figure}

Since the decay modes of $H^{\pm}$ including a neutralino happens along with also a chargino, we conclude this section by considering the chargino masses in the models as shown in Figure \ref{fig:charginos_Hpm} with plots in the $m_{\tilde{\chi}_{1}^{\pm}}-m_{H^{\pm}}$ and $m_{\tilde{\chi}_{2}^{\pm}}-m_{H^{\pm}}$ planes for CMSSM (left panel), NMSSM (middle panel) and UMSSM (right panel). The color coding is the same as Figure \ref{fig:m0m21_Hpm}. The solutions in the CMSSM framework are allowed by the constraints only when they yield $m_{\tilde{\chi}_{1}^{\pm}}\gtrsim 1$ TeV (seen from the black points). {Even though it is kinematically allowed ($m_{H^{\pm}}\sim m_{\tilde{\chi}_{1}^{0}}+m_{\tilde{\chi}_{1}^{\pm}}$), CMSSM may not provide significant decay processes in which the charged Higgs boson decays into a chargino and neutralino.} On the other hand, the same  constraints can allow lighter chargino solutions in NMSSM and UMSSM as $m_{\tilde{\chi}_{1}^{\pm}}\gtrsim 500$ GeV, while the heavier mass scales for the charged Higgs boson ($\gtrsim 10$ TeV) are also allowed. In this context, NMSSM and UMSSM may distinguish themselves from CMSSM, if they can yield significant $H^{\pm}\rightarrow \tilde{\chi}^{0}\tilde{\chi}^{\pm}$ processes. The bottom panels show also the second chargino may be effective in the charged Higgs decay modes, since the heavier charged Higgs boson is allowed and the constraints bound the second chargino mass from below as $m_{\chi^\pm_2}\gtrsim 2$ TeV in CMSSM.

\section{Production and Decay Modes of $H^{\pm}$}
\label{sec:res}

Direct production processes of the charged Higgs boson at the LHC is
rather difficult, since its production rate is proportional to the
Yukawa couplings of the Higgs bosons with the quarks from the first
two families. It is rather suppressed even when the charged Higgs
boson is light, because the Yukawa couplings associated with the
first two-family matter fields are quite small. On the other hand,
the charged Higgs bosons can be produced through the top-pair
productions if they are sufficiently light that the $t\rightarrow
H^{\pm}b$ process is kinematically allowed.  In such processes, the
charged Higgs boson can leave its marks through the
$H^{\pm}\rightarrow \tau^{\pm}b$ decay processes. In the cases with
heavy charged Higgs bosons, the charged Higgs boson is produced at
the LHC in association with either top-quark \cite{ATLAS:2016qiq} or
$W^{\pm}-$ boson \cite{BarrientosBendezu:1998gd}. The production
processes associated with top-quark do not provide clear signal due
to a large number of jets involved in the final states.
Nevertheless, those with $W^{\pm}$ boson might be expected to be
relatively clear signal, but such processes are suppressed by the
irreducible background processes from the top-pair production
processes \cite{Moretti:1998xq}. In this context, it is not easy to
detect the charged Higgs boson at the LHC. Even though an exclusion
limit is provided for the charged Higgs boson, it can be excluded up to
$m_{H^{\pm}} \sim 600$ GeV only for low $\tan\beta$ ($\sim 1$)
\cite{ATLAS:2016qiq}, which allows lighter charged Higgs boson
solutions when $\tan\beta$ is large.

\begin{table}[h!]
\centering
\scalebox{0.7}{
\begin{tabular}{|c|cc|cc|cc|}
\hline  &&&&&& \\
Observables & \multicolumn{2}{c|}{CMSSM} & \multicolumn{2}{c|}{NMSSM} & \multicolumn{2}{c|}{UMSSM} \\ \hline &&&&&& \\
 &  $m_{H^{\pm}}$(GeV) & $\sigma$(pb) & $m_{H^{\pm}}$(GeV) & $\sigma$(pb) & $m_{H^{\pm}}$(GeV) & $\sigma$(pb) \\ \hline &&&&&& \\ & 2019 & $4.5\times 10^{-5}$ & 1011 & $1.0\times 10^{-3}$ & 551 & $1.6\times 10^{-2}$ \\ &&&&&& \\ $pp\rightarrow tH^{\pm}$ & 3001 & $3.1\times 10^{-6}$ & 2055 & $1.2\times 10^{-4}$ & 1015 & $8.3\times 10^{-4}$ \\ &&&&&& \\
 & 4002 & $1.0\times 10^{-7}$ & 5849 & $5.8\times 10^{-9}$ & 2061 & $1.7\times 10^{-5}$ \\ \hline &&&&&& \\
 & 2019 & $5.2\times 10^{-6}$ & 1011 & $1.3\times 10^{-4}$ & 551 & $1.0\times 10^{-3}$ \\ &&&&&& \\
 $pp\rightarrow W^{\mp}H^{\pm}$ & 3001 & $4.2\times 10^{-7}$ & 2055 & $1.8\times 10^{-5}$ & 1015 & $6.4\times 10^{-5}$ \\ &&&&&& \\
 & 4002 & $1.7\times 10^{-8}$ & 5849 & $1.3\times 10^{-9}$ & 2061 & $2.0\times 10^{-6}$ \\ \hline &&&&&& \\ & 2019 & $4.8\times 10^{-8}$ & 1011 & $4.0\times 10^{-4}$ & 551 & $2.7\times 10^{-4}$ \\ &&&&&& \\ $pp\rightarrow H^{\mp}H^{\pm}$ & 3001 & $3.7\times 10^{-10}$ & 2055 & $5.9\times 10^{-6}$ & 1015 & $1.0\times 10^{-5}$ \\ &&&&&& \\
& 4002 & $1.5\times 10^{-12}$ & 5849 & $3.0\times 10^{-18}$ & 2061 & $4.0\times 10^{-8}$ \\ \hline &&&&&& \\
& 2019 & $1.6\times 10^{-8}$ & 1011 & $5.2\times 10^{-6}$ & 551 & $1.3\times 10^{-4}$ \\ &&&&&& \\
$pp\rightarrow H^{0}_{1,2,3}H^{\pm}$ & 3001 & $9.3\times 10^{-11}$ & 2055 & $1.5\times 10^{-8}$ & 1015 & $4.4\times 10^{-6}$ \\ &&&&&& \\
 & 4002 & $2.4\times 10^{-13}$ & 5849 & $1.8\times 10^{-15}$ & 2061 & $1.3\times 10^{-8}$ \\ \hline &&&&&& \\
& 2019 & $1.7\times 10^{-5}$ & 1011 & $4.1\times 10^{-2}$ & 551 & $7.2\times 10^{-3}$ \\ &&&&&& \\
$pp\rightarrow t\bar{b}H^{\pm}$ & 3001 & $1.4\times 10^{-6}$ & 2055 & $1.8\times 10^{-2}$ & 1015 & $3.4\times 10^{-4}$ \\ &&&&&& \\
& 4002 & $3.2\times 10^{-8}$ & 5849 & $1.7\times 10^{-9}$ & 2061 & $6.7\times 10^{-7}$ \\ \hline
\end{tabular}}
\caption{The charged Higgs boson production modes and cross-sections
over some benchmark points (we used the centre of mass energy
$\sqrt{s}$=14 TeV). The points have been chosen so as to be consistent
with all of the employed constraints discussed in Section
\ref{sec:scan}.} \label{table2}
\end{table}

Before considering the possible decay modes, we present the possible
production channels for the charged Higgs production over some
benchmark points in Table \ref{table2}. These values were obtained
after implementing our models into CalcHEP \cite{calchep}. The
points have been chosen so as to be consistent with the employed
constraints discussed in Section \ref{sec:scan}. The largest
contributions to the charged Higgs production come from the
processes involving with the top quark with the cross-section $\sim
10^{-5}$ pb in CMSSM, while this processes can reach up to $\sigma
\sim 10^{-3}$ pb in NMSSM and $\sigma \sim 10^{-2}$ pb in UMSSM.
Even though the improvement in the latter models is quite
significant, it is mostly because of the different mass scales of
the charged Higgs boson. As discussed in the previous section,
$m_{H^{\pm}} \gtrsim 2$ TeV is not allowed by the constraints in
CMSSM, $m_{H^{\pm}} \gtrsim 1$ TeV in NMSSM and $m_{H^{\pm}} \gtrsim
500$ GeV in UMSSM are allowed. However, considering
$\sigma(pp\rightarrow tH^{\pm})$, one should still note that NMSSM
still yields one magnitude larger cross-section ($\sim 10^{-4}$ pb)
in comparison to CMSSM and UMSSM for this production channel, if one
considers the similar mass scales for the charged Higgs boson
($m_{H^{\pm}} \sim 2$ TeV). Similar discussion holds for
$\sigma(pp\rightarrow t\bar{b}H^{\pm})$. The other channels can also
be seen in Table \ref{table2}, and as is stated before, they are
either negligible or a few magnitudes smaller than those involving
with top-quark.

Despite such small cross-sections in comparison to, for instance, the SM-like Higgs boson \cite{ATLAS:2017lpi}, some possible signal processes relevant to the charged Higgs boson become observable with a larger center of mass and luminosity. In addition, as stated before, the charged Higgs decays are crucial, since it does not exist in the SM, and these decays can
also be distinguishing between models. In this sense, we consider its decays and discuss the channels in a variety of SUSY models, which may play an important role in detecting the charged Higgs boson in future collider experiments. Depending on its mass, the charged Higgs boson can decay into either a pair of SUSY particles or the SM particles. Since the current LHC results imply rather a heavy mass spectrum for the squarks and gluinos, it might be possible to realize $H^{\pm}\rightarrow \tilde{t}\tilde{b} (\tilde{\tau}\tilde{\nu}_{\tau})$ processes which yield matter sparticles in their final states. However, these channels are hardly possible when the SUSY models are constrained from the GUT scale with the universal boundary conditions. However, it might still be possible that the charged Higgs boson can decay into a pair of chargino-neutralino. If SUSY particles are so heavy that the charged Higgs is not kinematically allowed to decay into the SUSY particles, then the SM particles take over and provide dominant decay channels. Since only the Yukawa couplings to the third family are significant, the final states are expected to include either third family quarks or leptons. The dominant decay channel is realized as $H^{\pm}\rightarrow tb$. Indeed, it is not surprising to realize the dominant channel as $H^{\pm}\rightarrow tb$ in all the cases when the charged Higgs boson is heavy, while, other decay channels can also be identified up to considerable percentage in some models. In this context, we start presenting our results for the $H^{\pm}\rightarrow tb$ process first, then we include other possible channels in our consideration.

\subsection{$H^{\pm}\rightarrow tb$}
\begin{figure}[ht!]\hspace{-1.0cm}
\includegraphics[scale=0.45]{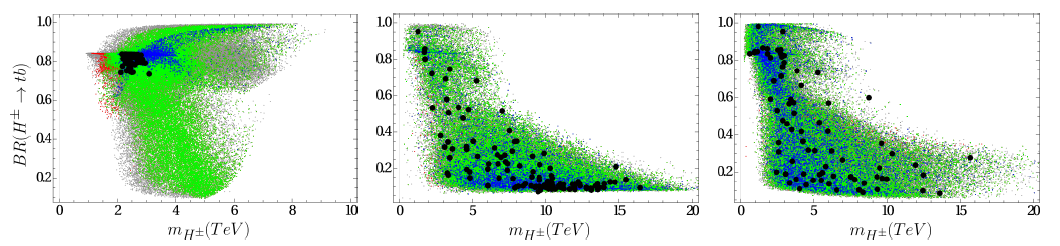}
\caption{The plots in the ${\rm BR}(H^{\pm}\rightarrow tb)-m_{H^{\pm}}$ plane for CMSSM (left), NMSSM (middle) and UMSSM (right). The color coding is the same as Figure \ref{fig:m0m21_Hpm}.}
\label{fig:Htb}
\end{figure}

Figure \ref{fig:Htb} represents our results for ${\rm
BR}(H^{\pm}\rightarrow tb)$ in correlation with $m_{H^{\pm}}$ in
CMSSM (left), NMSSM (middle) and UMSSM (right).  As mentioned before,
it provides the main decay channel for the charged Higgs boson in a
possible signal, which could be detected in future collider
experiments. When the DM constraints are applied (black points) top
of the LHC constraints, CMSSM allows this channel only up to $80\%$,
and it leaves a slight open window for the other possible decay
modes. The constraints also bound this process from below as  ${\rm
BR}(H^{\pm}\rightarrow tb) \gtrsim 70\%$. Hence, even though CMSSM
allows some other channels, their branching ratios cannot be larger
than $30\%$. In the case of NMSSM, it is possible to find solutions
in which the charged Higgs boson only decays into $tb$, {there is not
any lower bound provided by the constraints.} In other
words, it is possible to realize ${\rm BR}(H^{\pm}\rightarrow
tb)\sim 10\%$, which means one can identify some other channels as
the main channel, which are discussed later. Similar results can be
found also in the UMSSM framework. In this context, there is a wide
portion in the fundamental parameter space of NMSSM and UMSSM which
distinguishes these models from CMSSM.

\subsection{$H^{\pm}\rightarrow \tau \nu_{\tau}$}

\begin{figure}[ht!]\hspace{-1.0cm}
\includegraphics[scale=0.45]{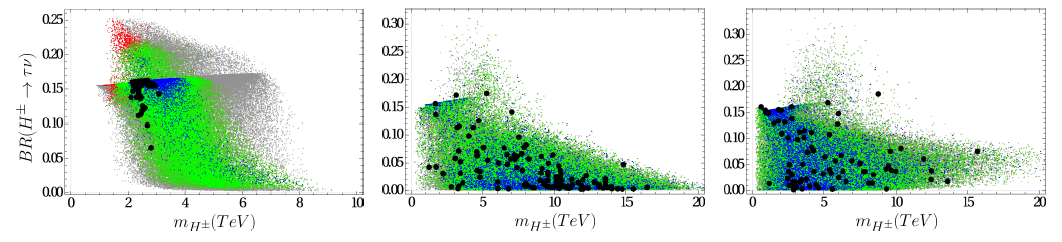}
\caption{The plots in the ${\rm BR}(H^{\pm}\rightarrow \tau\nu)-m_{H^{\pm}}$ plane for CMSSM (left), NMSSM (middle) and UMSSM (right). The color coding is the same as Figure \ref{fig:m0m21_Hpm}.}
\label{fig:Htaunu}
\end{figure}

Figure \ref{fig:Htaunu} displays our result for the ${\rm
BR}(H^{\pm}\rightarrow \tau\nu)$ in correlation with $m_{H^{\pm}}$
in CMSSM (left), NMSSM (middle) and UMSSM (right). The color coding
is the same as Figure \ref{fig:m0m21_Hpm}. All three models allow
this channel only up to about $20\%$. This channel is expected to be
dominant when $H^{\pm}\rightarrow tb$ is not allowed, i.e.
$m_{H^{\pm}} < m_{t}+m_{b}$ \cite{CMS:2016szv}. Even though there is
not {a} certain constraint through this leptonic decay of the charged
Higgs, and it can provide relatively clearer signal and less
uncertainty, it {may not display} a possible signal and
distinguish the models through these leptonic processes.

\subsection{$H^{\pm}\rightarrow \tilde{\chi}_{i}^{\pm}\tilde{\chi}_{j}^{0}$}
\begin{figure}[ht!]\hspace{-1.0cm}
\includegraphics[scale=0.45]{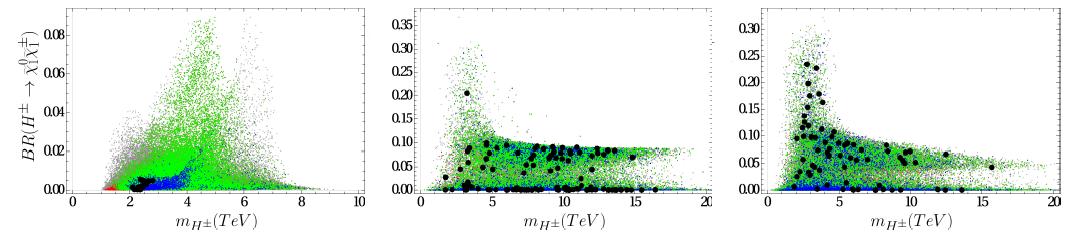}
\caption{The plots in the ${\rm BR}(H^{\pm}\rightarrow \tilde{\chi}_{1}^{0}\tilde{\chi}_{1}^{\pm})-m_{H^{\pm}}$ plane for CMSSM (left), NMSSM (middle) and UMSSM (right). The color coding is the same as Figure \ref{fig:m0m21_Hpm}.}
\label{fig:Hchi1cha1}
\end{figure}

Figure \ref{fig:Hchi1cha1}  shows the results for the ${\rm BR}(H^{\pm}\rightarrow \tilde{\chi}_{1}^{0}\tilde{\chi}_{1}^{\pm})$ in correlation with $m_{H^{\pm}}$ in CMSSM (left), NMSSM (middle) and UMSSM (right). The color coding is the same as Figure \ref{fig:m0m21_Hpm}. Even though it is possible to realize this process up to about $8\%$ (green) {in CMSSM}, the LHC measurements for the SM-like Higgs boson (blue) bound it as ${\rm BR}(H^{\pm}\rightarrow \tilde{\chi}_{1}^{0}\tilde{\chi}_{1}^{\pm})
\gtrsim 2\%$. However, these solutions do not satisfy the WMAP and
Planck bounds on the relic abundance of the LSP neutralino. When one
employs the DM constraints (black) it is seen that ${\rm
BR}(H^{\pm}\rightarrow
\tilde{\chi}_{1}^{0}\tilde{\chi}_{1}^{\pm})\lesssim 1\%$. Hence, a
possible signal involving with the
$H^{\pm}\tilde{\chi}_{1}^{0}\tilde{\chi}_{1}^{\pm}$ process {is hardly} realized in the CMSSM framework, while it is open in the NMSSM and
UMSSM frameworks up to about $20\% - 25\%$ consistently with all the
constraints including the DM ones. Note that even though the solutions {presented} in Figure \ref{fig:Hchi1cha1} are enough to claim a sensible difference from CMSSM, better results for the branching ratios in NMSSM and UMSSM may still be obtained with more thorough statistics.

As discussed in Section \ref{sec:spectrum}, the heavier neutralino
and chargino mass eigenstates are allowed to participate in
$H^{\pm}\rightarrow \tilde{\chi}_{i}^{0}\tilde{\chi}_{j}^{\pm}$.
However, the heavier ones continue to decay into the lighter mass
eigenstates, and each decay cascade gives a suppression unless their
branching ratio is large (${\rm BR}\sim 100\%$). In this context,
even though their signal is not as strong as $H^{\pm}\rightarrow
\tilde{\chi}_{1}^{0}\tilde{\chi}_{1}^{\pm}$, their contributions
would be at the order of minor contributions in comparison to ${\rm
BR}(H^{\pm}\rightarrow \tilde{\chi}_{i}^{0}\tilde{\chi}_{j}^{\pm})$.

\begin{table}[ht!]
\centering
\scalebox{0.65}{
\begin{tabular}{|c|cc|cc|cc|}
\hline &&&&&& \\ Parameters & \multicolumn{2}{c|}{CMSSM} & \multicolumn{2}{c|}{NMSSM} & \multicolumn{2}{c|}{UMSSM} \\ \hline &&&&&& \\ & Min(\%) & Max(\%) & Min(\%) & Max(\%) & Min(\%) & Max(\%) \\ \hline &&&&&&\\
BR($H^{\pm}\rightarrow \tilde{\chi}_{1}^{0}\tilde{\chi}_{1}^{\pm}$) & $-$ & 0.5 & $-$ & 20 & $-$ & 23 \\ &&&&&& \\
BR($H^{\pm}\rightarrow \tilde{\chi}_{2}^{0}\tilde{\chi}_{1}^{\pm}$) & $-$ & $-$ & $-$ & 3 & $-$ & 1 \\ &&&&&& \\
BR($H^{\pm}\rightarrow \tilde{\chi}_{3}^{0}\tilde{\chi}_{1}^{\pm}$) & $-$ & - & $-$ & 24 & $-$ & 21 \\ &&&&&& \\
BR($H^{\pm}\rightarrow \tilde{\chi}_{4}^{0}\tilde{\chi}_{1}^{\pm}$) & $-$ & - & $-$ & 26 & $-$ & 25 \\ &&&&&& \\
BR($H^{\pm}\rightarrow \tilde{\chi}_{5}^{0}\tilde{\chi}_{1}^{\pm}$) & $-$ & $-$ & $-$ & 25 & $-$ & 19 \\ &&&&&& \\
BR($H^{\pm}\rightarrow \tilde{\chi}_{6}^{0}\tilde{\chi}_{1}^{\pm}$) & $-$ & $-$ & $-$ & $-$ & $-$ & 8 \\ &&&&&& \\
BR($H^{\pm}\rightarrow \tilde{\tau}\tilde{\nu}$) & $-$ & 13 & $-$ & 33 & $-$ & 5 \\ &&&&&& \\
BR($H^{\pm}\rightarrow \tilde{t}\tilde{b}$) & $-$ & $-$ & $-$ & 35 & $-$ & 8 \\ &&&&&& \\
BR($H^{\pm}\rightarrow A^{0}_{1}W^{\pm}$) & $-$ & $-$ & $-$ & 43 & $-$ & $-$ \\ &&&&&& \\
BR($H^{\pm}\rightarrow H^{0}_{2}W^{\pm}$) & $-$ & $-$ & $-$ & 16 & $-$ & 2 \\ &&&&&& \\
BR($H^{\pm}\rightarrow ZW^{\pm}$) & $-$ & $-$ & $-$ & 3 & $-$ & 2 \\ &&&&&& \\
BR($H^{\pm}\rightarrow tb$) & 73 & 83 & 7 & 95 & 8 & 98 \\ &&&&&& \\
BR($H^{\pm}\rightarrow \tau \nu$) & - & 16 & $-$ & 17 & $-$ & 18\\
\hline
\end{tabular}}
\caption{Minimum and maximum rates for various decay modes of the
Charged Higgs boson obtained in the parameter scan. The values have
been chosen so as to be consistent with all the constraints
applied.} \label{table3}
\end{table}

In addition to the charginos and neutralinos, the Higgs boson can be
allowed to decay into some other supersymmetric particles, which
could be, in principle, a pair of $\tilde{t}\tilde{b}$ or
$\tilde{\tau}\tilde{\nu}$. As shown in Figure
\ref{fig:stopsbottom_Hpm}, the $H^{\pm}\rightarrow
\tilde{t}\tilde{b}$ process is not kinematically allowed in CMSSM,
while it is open in NMSSM and UMSSM. However, the large top-quark
background significantly suppresses such processes. On the other
hand, the $H^{\pm}\rightarrow \tilde{\tau}\tilde{\nu}$ process is
possible in all models. Moreover, since it happens through the
leptonic processes, the signal could be clear for such decay
processes. The minimum and maximum rates for various decay modes of
the Charged Higgs boson obtained in the parameter scan are
represented in Table \ref{table3}. The values have been chosen as to
be consistent with all the constraints applied. As mentioned before,
the dominant decay channel, $H^{\pm}\rightarrow t\bar{b}$,
does not leave {too much space} for the other modes in CMSSM,
while it is possible to realize this decay mode as low as a few
percent, and these models can significantly yield other decay modes
such as those with neutralino and chargino, stop and bottom, and/or
stau and neutralino, which can be as high as about 30$\%$ in NMSSM
and UMSSM. In addition, the decay processes including other Higgs
bosons can be also available in the latter models. For instance,
NMSSM allows the process, $H^{\pm}\rightarrow A^{0}_{1}W^{\pm}$ up
to $43\%$,  while the $H^{\pm}\rightarrow H^{0}_{2}W^{\pm}$ process can
be also realized up to $16\%$. The latter process is also allowed in
UMSSM up to about $30\%$.



\section{Conclusion}
\label{sec:conc} We perform numerical analyses for the CMSSM, NMSSM and UMSSM to probe the allowed mass ranges
for the charged Higgs boson and its possible decay modes as well as showing the allowed parameter spaces of these models. 
Since there is no charged scalar in SM, the charged Higgs boson can signal
the new physics as well as being distinguishable among the models
beyond SM. {Throughout {our} analyses, we find that it is possible to realize much heavier scales ($\gtrsim 10 $ TeV) in the NMSSM and UMSSM framework.}
In addition to the charged Higgs boson, we find
$m_{\tilde{t}}\gtrsim 2$ TeV in CMSSM, while it can be as light as
about 1 TeV in NMSSM and 500 GeV in UMSSM. These bounds on the stop
mass  mostly arise from the 125 GeV Higgs boson mass constraint,
while this constraint is rather relaxed in NMSSM and UMSSM because
of extra contributions from the new sectors in these models.
Besides, the sbottom mass cannot be lighter than about 2 TeV in CMSSM
and 1 TeV in  NMSSM and UMSSM. Such masses for the stop and sbottom
exclude $H^{\pm}\rightarrow \tilde{t}\tilde{b}$ in CMSSM, while it
can still be open in NMSSM and UMSSM. Another pair of supersymmetric
particles relevant to the charged Higgs boson decay modes is
$\tilde{\tau}$ and $\tilde{\nu}$, whose masses are bounded as
$m_{\tilde{\tau}} \gtrsim 500$ GeV and $m_{\tilde{\nu}}\gtrsim 1$
TeV. Even though their total mass is close by the charged Higgs
boson mass, there might be a small window which allows
$H^{\pm}\rightarrow \tilde{\tau}\tilde{\nu}$.  We also present the
masses for the chargino and neutralino, since the charged Higgs
boson can, in principle, decay into them.

While the heavy mass scales in NMSSM and UMSSM open more decay modes
for the charged Higgs boson, its heavy mass might be problematic in
its production processes. We list the possible production channels
and their ranges. The production channels, in which the charged
Higgs boson is produced associated with top-quark, provide the most
promising channels. Considering the same mass scale ($m_{H^{\pm}}
\sim 2$ TeV) in all the models CMSSM and UMSSM predict
$\sigma(pp\rightarrow tH^{\pm})\sim 10^{-5}$ pb, while NMSSM
prediction is one magnitude larger ($\sim 10^{-4}$ pb).  Similarly
UMSSM prediction fades away as $\sigma(pp\rightarrow
t\bar{b}H^{\pm}) \sim 10^{-7}$ pb, which is much lower than the CMSSM
prediction ($\sim 10^{-5}$ pb), while NMSSM yields
$\sigma(pp\rightarrow t\bar{b}H^{\pm}) \sim 10^{-2}$ pb. Even though
NMSSM predictions come forward in the charged Higgs prodictions,
UMSSM allows lighter charged Higgs mass solutions ($m_{H^{\pm}}\sim
500$ GeV) as well, and such solutions yield much larger production
cross-section as  $\sigma(pp\rightarrow tH^{\pm})\sim 10^{-2}$ pb
and $\sigma(pp\rightarrow t\bar{b}H^{\pm}) \sim 10^{-3}$ pb.

Even if these predictions for the charged Higgs boson production are
low in comparison to the SM-like Higgs boson, it can be attainable
with larger center of mass energy and luminosity. In addition, its
decay modes are completely distinguishable from any neutral scalar,
and hence it can manifest itself through some decay processes. The
dominant decay mode for the charged Higgs boson in CMSSM is mostly to
$tb$ with $70\% \lesssim {\rm BR}(H^{\pm}\rightarrow tb) \lesssim
80\%$, while it is also possible to realize $H^{\pm}\rightarrow
\tau\nu$ and $H^{\pm}\rightarrow \tilde{\tau}\tilde{\nu}$ up to
about $20\%$. On the other hand, the allowed heavy mass scales in
NMSSM and UMSSM allow the modes $H^{\pm}\rightarrow
\tilde{t}\tilde{b},\tilde{\tau}\tilde{\nu},\tilde{\chi}_{i}^{\pm}\tilde{\chi}_{j}^{0}$
in addition to those realized in the CMSSM framework. Among these
modes, even though $\tilde{t}\tilde{b}$ channel distinguishes these
models from CMSSM, the large irreducible top-quark background can
suppress such processes; thus, it is not easy to probe the charged
Higgs boson through such a decay mode. Nevertheless, despite the
clear leptonic background, NMSSM and UMSSM imply similar predictions
for the $\tilde{\tau}\tilde{\nu}$ decay mode to the results from
CMSSM. This channel can probe the charged Higgs in future collider
experiments but it cannot distinguish the mentioned SUSY models. On
the other hand, $H^{\pm}\rightarrow
\tilde{\chi}_{i}^{\pm}\tilde{\chi}_{j}^{0}$ is excluded by the
current experimental constraints in the CMSSM framework, while it is
still possible to include this decay mode in NMSSM and UMSSM. Such
decay modes can be probed in the collider experiments through the
missing energy and CP-violation measurements. Additionally, the
lightest chargino mass in NMSSM and UMSSM is bounded from below as
$m_{\tilde{\chi}_{1}^{\pm}}\gtrsim 1$ TeV, which seems testable in
near future LHC experiments through the analyses for the
chargino-neutralino production processes.

\vspace{0.3cm}
\noindent \textbf{Acknowledgement}

This work were supported by Balikesir University
Research Projects Grant No BAP-2017/142 and BAP-2017/174. The work of L. SELBUZ is supported in
part by Ankara University-BAP under grant number 17B0443004.

\end{document}